\documentclass[%
reprint,
amsmath,amssymb,
aps,
]{revtex4-2}

\usepackage{graphicx}%
\usepackage{dcolumn}%
\usepackage{bm}%
\usepackage{hyperref}%

\usepackage{amsmath}	%
\usepackage{glossaries} %
\usepackage{color}
\usepackage{xspace}

\newcommand{\days}{\ensuremath{\mathrm{days}}\xspace}
\newcommand{\he}{HE 0435 -- 1223\xspace}

\setacronymstyle{long-short} %
\glsdisablehyper %
\newacronym{agn}{AGN}{active galactic nuclei}
\newacronym[plural={GPs}, longplural={Gaussian Processes}]{gp}{GP}{Gaussian Process}
\newacronym{ns}{NS}{Nested Sampling}
\newacronym{smc}{SMC}{Sequential Monte Carlo}
\newacronym{pt}{PT}{Parallel Tempering}
\newacronym{mcmc}{MCMC}{Markov chain Monte Carlo}

\begin{document}
	
\preprint{APS/123-QED}

\title{Global structure of the time delay likelihood}

\author{Namu Kroupa$^{1,2,3}$}
\email{nk544@cam.ac.uk}
\author{Will Handley$^{2,4}$}
\affiliation{$^1$Astrophysics Group, Cavendish Laboratory, J.J. Thomson Avenue, Cambridge, CB3 0HE, UK}
\affiliation{$^2$Kavli Institute for Cosmology, Madingley Road, Cambridge, CB3 0HA, UK} 
\affiliation{$^3$Engineering Laboratory, University of Cambridge, Cambridge CB2 1PZ, UK}
\affiliation{$^4$Institute of Astronomy, University of Cambridge, Madingley Road CB3 0HA, UK}

\date{\today}

\begin{abstract}
	
We identify a fundamental pathology in the likelihood for time delay inference which challenges standard inference methods.
By analysing the likelihood for time delay inference with Gaussian process light curve models, we show that it generically develops a boundary-driven ``W''-shape with a global maximum at the true delay and gradual rises towards the edges of the observation window. This arises because time delay estimation is intrinsically extrapolative. 
In practice, global samplers such as nested sampling are steered towards spurious edge modes unless strict convergence criteria are adopted. We demonstrate this with simulations and show that the effect strengthens with higher data density over a fixed time span. To ensure convergence, we provide concrete guidance, notably increasing the number of live points. Further, we show that methods implicitly favouring small delays, for example optimisers and local MCMC, induce a 
bias towards larger $H_0$. Our results clarify failure modes and offer practical remedies for robust fully Bayesian time delay inference.
	
\end{abstract}

\maketitle

\section{Introduction}

Time delay measurements constitute a primary observable for lensed-quasar time delay cosmography~\cite{refsdal1964possibility, treu, birrer, treu2022strong, birrer2025challenges} and reverberation mapping of active galactic nuclei~\cite{peterson1993reverberation, cackett2021reverberation, netzer1997reverberation}. In both settings the light curves are noisy, irregularly sampled and stochastic, yielding likelihood surfaces that are difficult to explore and that will strain standard pipelines as samples grow. The Vera C. Rubin Observatory’s Legacy Survey of Space and Time (LSST) is expected to increase the number of known lensed quasars by roughly two orders of magnitude~\cite{abell2009lsst, ivezic2019lsst} and ongoing spectroscopic programmes, including the SDSS‑V Black Hole Mapper, are expanding the number of reverberation mapping targets~\cite{kollmeier2019sdss, almeida2023eighteenth}. Automated, reliable inference of time delays with quantified uncertainties is therefore needed to control systematic errors in $H_0$ and to enable robust population studies.

A fully Bayesian framework is well suited to these problems as it represents uncertainty coherently and supports principled model comparison. Obtaining faithful posteriors, however, requires samplers that can navigate specific documented pathologies. Seasonal gaps and finite observing windows induce window-function aliases that generate multiple, well-separated modes in the delay posterior, often at separations comparable to the observing season~\cite{grier2017sloan, grier2019sloan, jennifer2019sloan, homayouni2020sloan, mcdougall2025litmus}. The likelihood also exhibits strong, nonlinear degeneracies between the delay and parameters controlling the intrinsic variability. In practice, this necessitates constraints on kernel flexibility in \gls{gp} regression or spline smoothness~\cite{hojjati2013robust, tewes2013cosmograil}. 
In reverberation mapping, the convolutional response leads to pronounced parameter degeneracies, so that the recovered lag covaries with the width and shape of the transfer function and with hyperparameters of the continuum variability, and the broad-line region geometry and kinematics exhibit strong covariances and frequent multimodality~\cite{zu2011alternative, li2013bayesian, pancoast2014modellingI, pancoast2014modellingII, starkey2017space}.
Historical cases highlight this point. For Q0957+561, early analyses supported distinct delays which a modern global treatment would reveal as multi-modality and were ultimately resolved by denser monitoring~\cite{vanderriest1989value, lehar1992radio, press1992time, pelt1995light, kundic1996robust}.
For B1608+656, degeneracies in the lens mass distribution were broken by combining time delays with high-resolution imaging and stellar kinematics in a unified Bayesian analysis which enabled evidence-based model comparison~\cite{suyu2009dissecting, suyu2010dissecting, koopmans2003hubble}.

Methods based on optimisation and locally exploring MCMC samplers frequently rely on favourable initialisation to terminate in the correct mode. In practice, visual matching and restricted re-initialisations impose implicit priors that limit exploration~\cite{dobler2015strong, liao2015strong, kumar2013cosmograil, courbin2017cosmograil}. 
Global samplers, including nested sampling, offer an alternative by sampling the prior subject to progressively tighter likelihood constraints. They can, with sufficiently many live points and appropriate proposals, recover well-separated modes and complex degeneracies, and provide the Bayesian evidence for model comparison by construction~\cite{skilling, ashton, buchner2023nested, kroupa2024kernel}. 
Although MCMC can estimate evidence through specialised procedures~\cite{chib1995marginal, gelman1998simulating, lartillot2006computing}, nested sampling delivers it natively. 
Nonetheless, the performance of any global method depends on configuration and diagnostics, and can degrade on highly structured posteriors of the kind encountered in time delay analyses.

Here we identify and analyse a further, intrinsic feature of the time delay likelihood that affects both local and global approaches. Under broad and commonly adopted assumptions, namely stationary \gls{gp} models for intrinsic variability, the likelihood as a function of delay exhibits a boundary-driven ``W''-shaped structure. As the overlap of shifted light curves shrinks near the window edges, inference often reduces to extrapolation, which systematically elevates the likelihood towards the boundaries. 
We show that while the boundary rise is generic, the central peak remains dominant under typical conditions, so the inference problem is well-posed in principle. The practical difficulty is algorithmic. Samplers must be configured to discover and correctly weight all modes in the presence of gradual boundary increases. Through numerical simulations, we demonstrate how this structure can mislead \gls{ns} unless care is taken with live-point budgets and proposal strategies. We quantify the consequence of the common tendency of optimisation and local samplers to bias delays to low values, which translates into a positive bias in $H_0$.

Beyond astrophysics, time delay estimation is a recurring statistical problem in underwater acoustics~\cite{urick1975principles}, wireless communications~\cite{turin2005introduction}, signal processing~\cite{chen2006time, gedalyahu2010time}, radar~\cite{quazi2003overview}, biology~\cite{heerah2021granger}, geophysics~\cite{schaff2005waveform}, finance~\cite{hoffmann2013estimation} and control applications~\cite{bjorklund2003review}.
From the perspective of sampling, \gls{ns} applied to the calculation of phase diagrams in materials science~\cite{partay2021nested} faces a similar computational challenge. There, \gls{ns} is used to generate samples from the Boltzmann distribution. One of the primary aims lies in the calculation of the critical temperature of a phase transition. Unbiased estimation thereof relies on proper sampling of an ordered low-entropy phase, corresponding to a narrow deep potential energy basin of, for example, a solid, and a disordered high-entropy phase, corresponding to a wide shallow basin of, for example, a liquid. 
Since \gls{ns} is initialised from the high-temperature disordered phase, discovering the solid phase can be computationally challenging~\cite{chew2023phase}.
Analogously, in the time delay problem, it is difficult to discover the true central mode while it is easy to sample the wider edge modes.

The paper is organised as follows. Section~\ref{sec:time-delayed-gaussian-processes} presents time delayed \glspl{gp}, Section~\ref{sec:toy-model} describes the toy model, Section~\ref{sec:synthetic-data-section} derives analytical results and Section~\ref{sec:numerical-investigation} contains a numerical investigation. Finally, Section~\ref{sec:conclusions} outlines the conclusions.

\section{Time delayed Gaussian Processes}~\label{sec:time-delayed-gaussian-processes}

Typically~\citep{rasmussen}, \glspl{gp} are defined as a collection of random variables, any finite subset of which has a joint Gaussian distribution. We take the alternative definition in weight-space here, in order to make the underlying parametric model explicit. That is, we view a \gls{gp} with mean function $m$ as linear regression with a possibly infinite number of basis functions $\phi_i$,
\begin{equation}\label{eqn:gp}
	f(t)=m(t)+\sum_{i=1}^\infty w_i \phi_i(t),
\end{equation}
where $w_i$ are the model parameters. The mean function $m$ and the basis $\phi_i$ may depend on further (hyper)parameters which we take at this stage as given. Placing a zero-mean Gaussian prior with diagonal covariance matrix on $w_i$ thus induces a prior in function space.

In the following, we will also be interested in the time delayed function,
\begin{equation}\label{eqn:time-delayed-gp}
	h(t)=f(t+\Delta t)=m(t+\Delta t)+\sum_{i=1}^\infty w_i\phi_i(t+\Delta t),
\end{equation}
for a given time delay $\Delta t$. Since the parameters $w_i$ are shared between the random variables $f(t)$ and $h(t')$, they are jointly distributed as a Gaussian. The mean directly follows from computing $\mathbb{E}[f(t)]$ and $\mathbb{E}[h(t')]$, while the covariance may be obtained similarly. For example, the off-diagonal entries are computed from $\mathbb{E}[(f(t)-\mathbb{E}f(t))(h(t')-\mathbb{E}h(t'))]$. Thus, integrating out the parameters leaves us with the marginal likelihood
\begin{align}\label{eqn:time-delayed-gp-first-eqn}
	\begin{bmatrix}
		f(t)\\
		h(t')
	\end{bmatrix}
	\sim 
	\mathcal{N}\Biggl(
	&\begin{bmatrix}
		m(t)\\
		m(t'+\Delta t)
	\end{bmatrix}, \\
	&\begin{bmatrix}
		k(t,t)&k(t,t'+\Delta t)\\
		k(t+\Delta t, t')&k(t'+\Delta t,t'+\Delta t)
	\end{bmatrix}
	\Biggr) \nonumber
\end{align}
for given $t$, $t'$ and $\Delta t$. Here, we have defined the kernel
\begin{align}
	k(t,t')&=\sum_{i=1}^\infty\sum_{j=1}^\infty\mathbb{E}[w_iw_j]\phi_i(t)\phi_j(t')\\
	&=\sum_{i=1}^{\infty}\lambda_i^2\phi_i(t)\phi_i(t'),
\end{align}
where $\lambda_i^2$ is the prior variance of each parameter.
Specific choices of $\phi_i$ and corresponding priors on $w_i$ return common kernels such as Mat\'ern or squared exponential kernels.
The inductive bias of the model arises both from the choice of the basis and the rate of decay of the prior variance $\lambda^2_i$ as $i\rightarrow \infty$.
Different choices for these result in the function $f$ having different smoothness properties.
In practice, certain functional forms of the kernel are invoked which guarantee the existence of such an expansion in basis functions and the kernel hyperparameters then control $\phi_i$ and $\lambda_i$, so that one only has to work with Equation~\ref{eqn:time-delayed-gp-first-eqn}.

As expected, infinitely many parameters render \glspl{gp} flexible models. In fact, if $\phi_i$ is a complete basis, it is in principle possible to re-expand the mean function $m$ in this basis and absorb it in the sum in Equation~\ref{eqn:gp}. It is therefore reasonable to ask whether the inclusion of a separate mean function is necessary in the first place.

The necessity arises from noting that, for a finite amount of data and infinitely many basis functions, the regularisation imposed by the prior can indeed be very strong, in the sense that the posterior $f$ is completely prior-dominated as we move away from a data point. This problem is exacerbated by the fact that the task of inferring the time delayed function $h$ is intrinsically extrapolative. That is, if the time delay $\Delta t$ is large, $h$ is far away from $f$ along the $t$-axis so that $h$ and $f$ are effectively uncorrelated.

To be more precise, suppose we are given noise-free data ${y_1(t_1),\dots,y_1(t_{n_\mathrm{data}})}$ and ${y_2(t_1),\dots,y_2(t_{n_\mathrm{data}})}$ for $f$ and its time-shifted copy $h$, respectively, observed at the times $t_1,\dots,t_{n_\mathrm{data}}$. To aid interpretation, the posterior $p(\Delta t\mid y_1,y_2)$ of the time delay may be written as (Appendix~\ref{appendix:derivation-time-delay-posterior})
\begin{equation}\label{eqn:time-delay-posterior}
	p(\Delta t\mid y_1,y_2)\propto \int\mathrm{d}\mathbf{w}\, p(y_2\mid \Delta t,\mathbf{w})p(\mathbf{w}\mid y_1),
\end{equation}
where we assumed a uniform prior on~$\Delta t$ and the integration is performed over all parameters~$\mathbf{w}$. This states that we first compute the posterior $p(\mathbf{w}\mid y_1)$ of the parameters given only light curve~$y_1$. We then draw a set of parameters $\mathbf{w}$ from this posterior, which completely determine the function $f$. Given the time delay $\Delta t$, we now have full information to construct the time-shifted copy $h$ (by simply shifting $f$ along the $t$-axis), which allows us to evaluate the likelihood $p(y_2\mid \Delta t,\mathbf{w})$ at the second light curve $y_2$.
Since the time-shift implies that we do not evaluate $f$ at the times $t_1,\dots,t_{n_\mathrm{data}}$ but instead at $t_1+\Delta t,\dots,t_{n_\mathrm{data}}+\Delta t$, extrapolation and interpolation are required. The amount of extrapolation depends on~$\Delta t$.

In general, the extrapolated function averaged over $\mathbf{w}$, which is the posterior mean function, is a linear combination of kernel functions centred on each data point~\citep{rasmussen}. For a stationary \gls{gp}, the kernels are localised, i.e. the extrapolation effectively does not depend on the data beyond a certain decorrelation length scale. Instead, the \gls{gp} returns to the prior, as previously claimed. We will see in Section~\ref{sec:synthetic-data-section} that the lack of extrapolative capability of a \gls{gp} also renders the structure of the time delay posterior (Equation~\ref{eqn:time-delay-posterior}) pathological, causing additional practical problems.

Because time delay inference necessarily compares shifted light curves beyond their overlap window, extrapolation is unavoidable. This necessitates adding non-stationary structure, for example via non-stationary kernels or explicit mean functions, so that trends can be represented. Here, we intentionally restrict the investigation to the simplest class, namely a stationary exponential kernel and a constant mean, so that the likelihood structure and its boundary effects can be isolated and analysed cleanly. The mechanism we uncover is model-agnostic in the sense that it is driven by finite-window overlap rather than the particular functional prior. Extensions to more flexible models are left to future work.

\section{Toy model}\label{sec:toy-model}

In this section, we analyse the posterior of the time delay by considering the following synthetic data setting. 
We consider two light curves, $\mathbf{y}_1$ and $\mathbf{y}_2$, both sampled from a \gls{gp} with zero mean and an exponential kernel, 
\begin{equation}
	k(t,t')=A^2\exp\left(-\frac{|t-t'|}{\ell}\right),
\end{equation}
as this corresponds to the damped random walk model of quasar variability~\cite{macleod2010modeling, zu}.
The known hyperparameters are ${\bm{\theta}=(A,\ell, \sigma, \Delta t)}$, with the amplitude $A$, length scale $\ell$, noise $\sigma$ and time delay $\Delta t$, respectively.
The observation times $\mathbf{t}$ are chosen to be $n_\mathrm{data}$ equally spaced points between some minimum and maximum observation time, which we set to $T_\mathrm{min}=0\,\days$ and $T_\mathrm{max}=10^3\,\days$, respectively, as this reflects realistic total observation times.
We set $A=1$ in units of magnitude without loss of generality, as any other value may be reduced to this choice by appropriate rescaling. Furthermore, we set the noise to be negligible in comparison to the signal, $\sigma=10^{-2}\ll A$. Thus, the data are a sample from:
\begin{equation}\label{eqn:time-delayed-GP-pair}
	\begin{bmatrix}\mathbf{y}_1\\\mathbf{y}_2\end{bmatrix}\sim \mathcal{N}
	\left(
	\mathbf{0},
	\begin{bmatrix}
		\mathbf{K}(\mathbf{t},\mathbf{t})+\sigma^2\mathbf{I}&\mathbf{K}(\mathbf{t},\mathbf{t}-\Delta t\mathbf{1})\\
		\mathbf{K}(\mathbf{t}-\Delta t\mathbf{1},\mathbf{t})&\mathbf{K}(\mathbf{t},\mathbf{t})+\sigma^2\mathbf{I}
	\end{bmatrix}
	\right),
\end{equation}
where the covariance matrix in each block is given by ${[\mathbf{K}(\mathbf{t},\mathbf{t})]_{ij}=k(t_i,t_j)}$ and $\mathbf{I}$ is the identity matrix.
Given such data, we would in practice sample parameters using the likelihood
\begin{equation}\label{eqn:synthetic-data-likelihood}
	L(\bm{\theta})=\frac{1}{\sqrt{|2\pi \mathbf{K}_{\bm{\theta}}|}}\exp\left(-\frac12 \mathbf{y}^\top \mathbf{K}_{\bm{\theta}}^{-1} \mathbf{y}\right),
\end{equation}
where $\mathbf{y}_1$ and $\mathbf{y}_2$ are concatenated into $\mathbf{y}$ and $\mathbf{K}_{\bm{\theta}}$ is the full covariance matrix in Equation~\ref{eqn:time-delayed-GP-pair}, parameterised by the hyperparameters $\bm{\theta}$.  
Furthermore, uniform priors are set on all \gls{gp} hyperparameters. In general, we observe~\cite{kroupa2026quasars} that for real data sets, the bulk of the posterior mass is well contained within the chosen prior ranges, except for the length scale of the exponential kernel, which we attribute to the data (Appendix~\ref{appendix:weakly-determined-length-scales}).

\section{Analytical results}\label{sec:synthetic-data-section}

In this section, we use the toy model to investigate the log-likelihood structure of the time delay. The advantage in this setting is that we know the exact data-generating distribution and the values of the true (hyper)parameters. By applying the analytical understanding, we then show that the pathological likelihood structure may in principle be removed by a suitable prior.

\subsection{The log-likelihood structure}\label{sec:log-likelihood-structure}

In order to make statements about $L$ for any data sample, we consider the data-averaged log-likelihood and its variance, respectively,
\begin{align}
	\mathbb{E}_{\mathbf{y}}\log L(\bm{\theta})&=-\frac12 \mathrm{Tr}\left(\mathbf{K}_{\bm{\theta}}^{-1}\mathbf{K}\right)-\frac12 \log |2\pi \mathbf{K}_{\bm{\theta}}|, \label{eqn:data-averaged-log-likelihood}
	\\
	\mathbb{V}_{\mathbf{y}}\log L(\bm{\theta})&=\frac14 \sum_{ij}(\mathbf{K}^{-1}_{\bm{\theta}})_{ij}\left[\mathbf{K}_{ii}\mathbf{K}_{jj}+(\mathbf{K}_{ji})^2\right],
\end{align}
where $\mathbf{K}=\mathbb{E}_\mathbf{y}\mathbf{y}\mathbf{y}^\top$ is the true covariance of the data, i.e. the covariance matrix in Equation~\ref{eqn:time-delayed-GP-pair}. The first term is typically interpreted as representing the goodness of fit, whereas the second term penalises model complexity~\citep{rasmussen}. Note that this decomposition into goodness of fit and model complexity penalty is distinct from Occam's razor equation, $\log Z=\langle \log L\rangle_P - D_\mathrm{KL}$~\cite{hergt}, where $Z$ is the Bayesian evidence, $\langle \log L\rangle_P$ the posterior-averaged log-likelihood and $D_\mathrm{KL}$ the Kullback-Leibler divergence.

\begin{figure*}[!]
	\centering
	\includegraphics{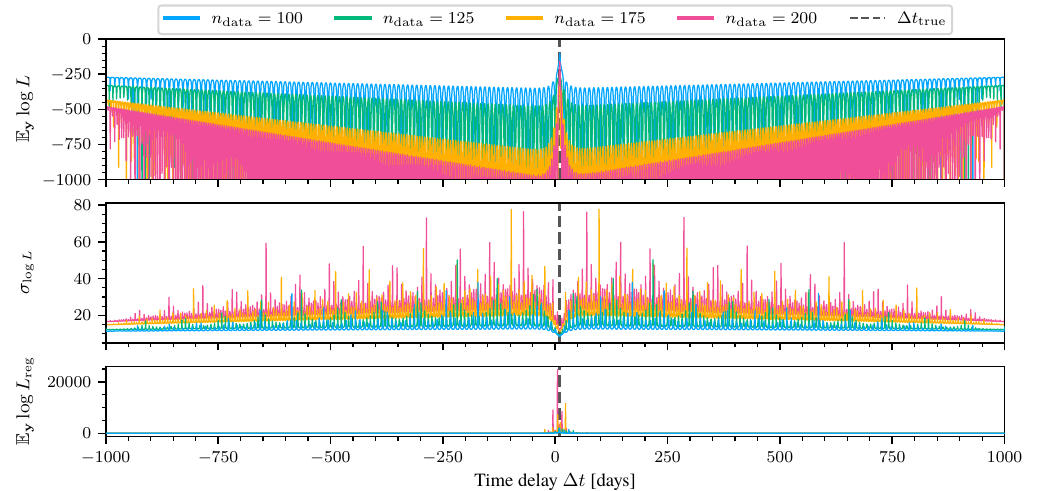}
	\caption{\textbf{Top:} Data-averaged log-likelihood $\mathbb{E}_\mathbf{y}\log L$ against the time delay $\Delta t$. The true time delay for this figure is ${\Delta t_{\mathrm{true}}=10\,\days}$. The log-likelihood is locally highly oscillatory and globally follows a ``W''-shape with a global maximum at the true time delay. \textbf{Middle:} The standard deviation $\sigma_{\log L}=\sqrt{\mathbb{V}_\mathbf{y}\log L}$ is negligible compared to the expectation. \textbf{Bottom:} Data-averaged regularised log-likelihood $\mathbb{E}_\mathbf{y}\log L_{\mathrm{reg}}$ formed by subtracting the trend favouring large time delays. Only the maxima in the vicinity of the true time delay remain.}
	\label{fig:data-averaged-loglikelihood}
\end{figure*}

The top subplot of Figure~\ref{fig:data-averaged-loglikelihood} shows $\mathbb{E}_\mathbf{y}\log L$. While $\mathbb{E}_\mathbf{y}\log L$ is highly oscillatory, it has a global maximum at the true time delay. 
Away from this peak, the log-likelihood gradually rises to local maxima at the edges of the time delay prior. Increasing the number of data points both sharpens the peak at the true time delay and the local maxima at the edges.
As stated previously, $\mathbb{E}_\mathbf{y}\log L$ lower bounds the log of the true likelihood, so that we cannot necessarily deduce whether the peak at the true time delay is dominant.
However, while the variance 
spikes at particular values of $\Delta t$, as shown in the middle subplot, it is overall negligible and $\mathbb{E}_\mathbf{y}\log L$ should be representative of an individual data set sample. 
As we will show in Section~\ref{sec:numerical-problems}, the gradual increase towards large time delays causes \gls{ns} to falsely sample large time delays. 

\begin{figure*}[!]
	\centering
	\includegraphics{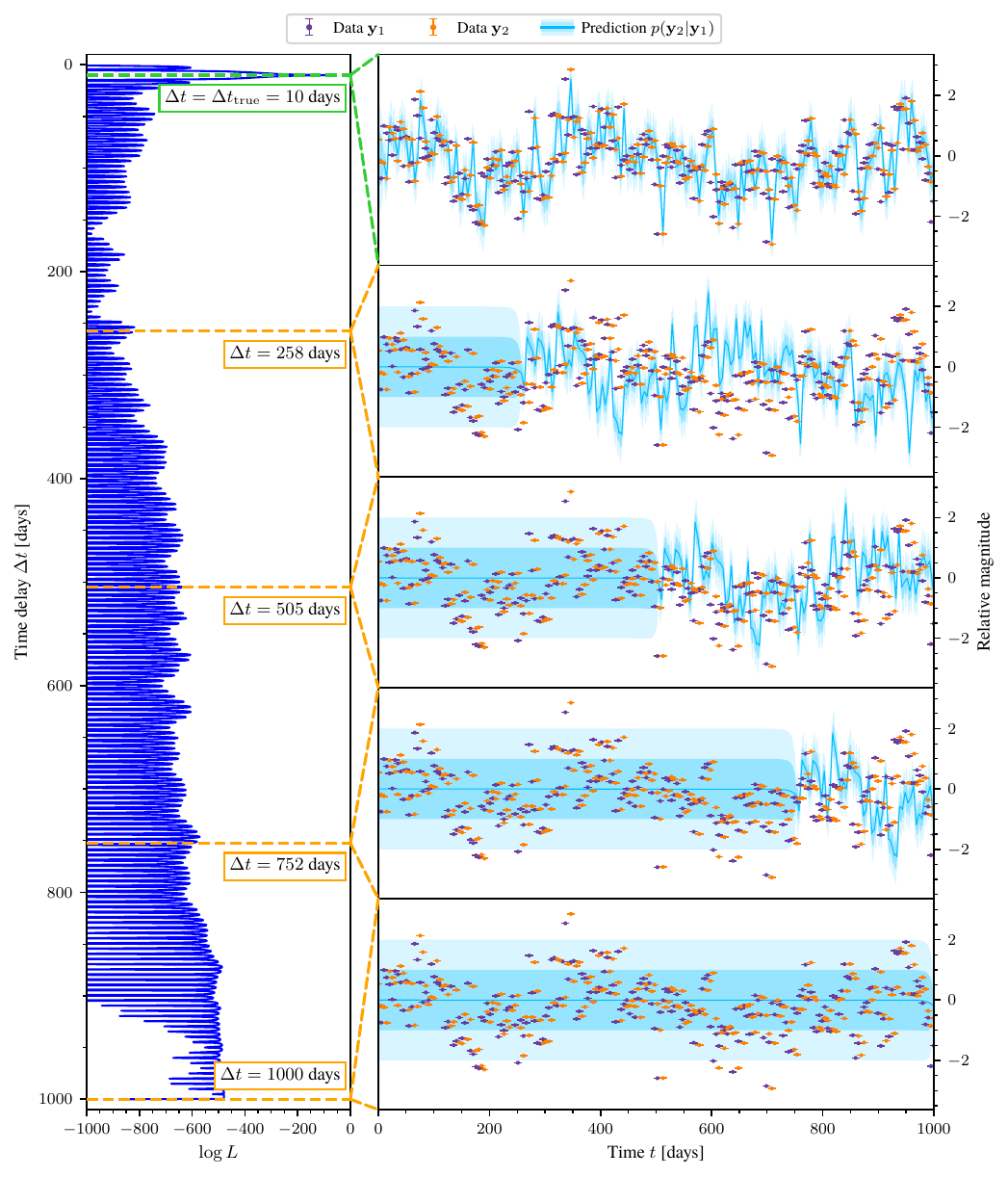}
	\caption{Illustration of the gradual increase in $\log L$ as the time delay $\Delta t$ increases. Two synthetic data sets shown on the right are sampled from the distribution defined in Equation \ref{eqn:time-delayed-GP-pair} with true time delay $\Delta t=10\,\days$. Subsequently, the log-likelihood $\log L$ of the \gls{gp} model is evaluated as a function of the time delay, shown on the left. The evaluation of $\log L$ is reformulated as visually comparing the predictive distribution $p(\mathbf{y}_2|\mathbf{y}_1)$ conditioned on $\mathbf{y}_1$ with the data set $\mathbf{y}_2$. This is shown in the sequence of plots on the right for different time delays. At the true time delay, the predictive distribution evidently matches $\mathbf{y}_2$. As the time delay is increased, the mean of $p(\mathbf{y}_2|\mathbf{y}_1)$ does not pass through the data $\mathbf{y}_2$ and $\log L$ decreases significantly. However, the \gls{gp} increasingly treats the non-overlapping data as noise, which provides a good fit. As $\Delta t$ is increased further, more data are fit as noise and $\log L$ rises gradually, leading to the ``W''-shape of $\log L$ (Figure~\ref{fig:data-averaged-loglikelihood}).}
	\label{fig:likelihood-illustration}
\end{figure*}

What causes this gradual increase in log-likelihood? 
As we show in Appendix~\ref{appendix:refined-argument-log-likelihood}, the log-likelihood at the true time delay is much larger than the log-likelihood at the edges.
Further, we show that the dominant contribution to the log-likelihood is given by the complexity penalty term. 
This renders large time delays unfavourable compared to the true time delay. 
In this sense, the likelihood is not pathological and the inference problem is well-specified. Namely, if we could sample perfectly from the posterior, we would always recover the true time delay. The difficulty of recovering the true time delay is therefore a problem with the method of sampling, more specifically a problem with the internal dynamics of the global sampling method induced by the gradual likelihood increase.

As shown in Appendix~\ref{appendix:refined-argument-log-likelihood}, the `$\chi^2$', representing the goodness of fit, enters as a subdominant contribution and, in expectation, is of similar size for both the true and large time delays. 
Any account of the gradual drift would therefore enter as an even lower-order contribution. 

From an intuitive standpoint however, the drift can be understood as follows (Figure~\ref{fig:likelihood-illustration}). Following the approach in Appendix~\ref{appendix:refined-argument-log-likelihood}, we first decompose the likelihood as
\begin{equation}
	p(\mathbf{y}_1,\mathbf{y}_2|\Delta t)=p(\mathbf{y}_2|\mathbf{y}_1,\Delta t)p(\mathbf{y}_1).
\end{equation}
The term $p(\mathbf{y}_1)$ is independent of $\Delta t$ and can be ignored. Only the predictive distribution $p(\mathbf{y}_2|\mathbf{y}_1,\Delta t)$ depends on the time delay. Thus, evaluating the likelihood is equivalent to asking how well $\mathbf{y}_1$ predicts $\mathbf{y}_2$.

We now imagine first fitting the light curve $\mathbf{y}_1$ with a \gls{gp}. This fit will interpolate the data well since we are using the correct \gls{gp} model which also generated the data. However, outside of the domain of the data, the \gls{gp} reverts to the prior and we effectively see a zero mean function with error bars $\pm A$. As we start increasing $\Delta t$ and 
monitor the overlap of the prediction $p(\mathbf{y}_2|\mathbf{y}_1,\Delta t)$ with $\mathbf{y}_2$,
we will see very good agreement 
at the true time delay, corresponding to the true mode in the posterior. 
As~$\Delta t$ increases
further, the match with the interpolation
clearly becomes worse. 
However, the extrapolated part with the wide error bars starts to overlap with $\mathbf{y}_2$ on one end. In this region, the fit does agree with the data, as all data points lie within the error bars of the predictive distribution. 
As $\Delta t$ is increased, this region grows while the region with the bad fit shrinks.
Formulated differently, the number of data points ignored grows proportionally to $\Delta t$. Therefore, the likelihood increases approximately linearly with $\Delta t$, as seen in Figure~\ref{fig:data-averaged-loglikelihood} and~\ref{fig:likelihood-illustration}.

As we increase the number of data points for a fixed total observation time, the data density increases. Therefore, when $\Delta t$ is increased, the rate at which the fit ignores data points increases as well. Consequently, the slope of the linear increase also increases with the data density, as observed previously in Figure~\ref{fig:data-averaged-loglikelihood}.

\subsection{Constructing a counterterm prior}\label{sec:counterterm-prior}

Having understood the structure of the log-likelihood, we ask at this point whether we can design a prior which cancels out the gradual increase towards the edges.

Above, we have argued that the likelihood can effectively be thought of the superposition of two competing models, namely a narrow tall peak at $\Delta t=0$ and a broad shallow peak at $\Delta t=\pm t_\mathrm{range}$. Hence, we can try to isolate just the $\Delta t=0$ peak by subtracting an appropriate term. 

To construct this term, we change our perspective and imagine two fictitious light curves $\mathbf{y}_1'$ and $\mathbf{y}_2'$ which are decorrelated, i.e. have a true time delay of ${\Delta t_\mathrm{true}'= t_\mathrm{range}}$. In this case, the largest likelihood occurs when the predictive distribution conditioned on $\mathbf{y}_1'$ extrapolates to noise and $\mathbf{y}_2'$ resides in this noise band.
But the likelihood for our original light curves $\mathbf{y}_1$ and $\mathbf{y}_2$ must be the same at $\Delta t=t_\mathrm{range}$ because the noise band has lost its memory of the first light curve. 
Hence, we should expect perfect cancellation at $\Delta t=t_\mathrm{range}$ when we subtract the two likelihoods: ${\log L(t_\mathrm{range})-\log L'(t_\mathrm{range})=0}$, where $L'$ is the likelihood of the fictitious data.

As we decrease the time delay, the log-likelihood in our fictitious system should decrease because the region of the ``good'' fit, in which $\mathbf{y}_2'$ resides in the noise band, shrinks. This decrease in log-likelihood should be proportional to $\Delta t$ as we are losing data points which fit the prediction well. At least in the vicinity of $\Delta t=t_\mathrm{range}$, we can therefore expect that the decrease in $\log L$ and $\log L'$ is the same.

This discussion motivates the definition of the regularised log-likelihood,
\begin{equation}\label{eqn:regularised-log-likelihood}
	\log L_\mathrm{reg}(\Delta t)=\log L(\Delta t) - \mathbb{E}_{\mathbf{y}_{t_\mathrm{range}}}\log L(\Delta t).
\end{equation}
Here, we have essentially subtracted the likelihood of the fictitious decorrelated data from the likelihood of the real data.
That is, the subtracted term is given by Equation~\ref{eqn:data-averaged-log-likelihood} but where $\mathbf{K}$ uses $\Delta t=t_\mathrm{range}$ as the true time delay.
Note that we also average the subtracted term over the fictitious data, as otherwise we would have to select a particular fictitious data set.

To investigate this new likelihood, we average over the data in order to be able to make generic statements for different realisations of the data, as done previously.
Thus, we obtain the data-averaged regularised log-likelihood,
\begin{equation}
	\mathbb{E}_{\mathbf{y}}\log L_\mathrm{reg}(\Delta t)=\mathbb{E}_\mathbf{y}\log L(\Delta t) - \mathbb{E}_{\mathbf{y}_{t_\mathrm{range}}}\log L(\Delta t).
\end{equation}
This is shown in the bottom subplot in Figure~\ref{fig:data-averaged-loglikelihood}. As expected, the trend in the log-likelihood towards $\pm t_\mathrm{range}$ is removed, leaving just the peak near the true time delay. Notably, the entire gradual drift is subtracted, suggesting that the above motivating argument for the construction of the regularised likelihood holds for a larger range of~$\Delta t$ than the argument initially suggests.

It is now tempting to interpret the subtracted term in Equation~\ref{eqn:regularised-log-likelihood} as a prior on $\Delta t$ which removes the full separation of the light curves and therefore acts as a counterterm. 
However, such a prior can over-regularise the likelihood and result in time delays smaller in absolute value than those obtained with flat priors. 
This happens because this prior does not distinguish between the region around the true peak, where an unbiased uniform prior is desirable, and large time delays, where prior constraints may be applicable. 
Equivalently, the shape of the prior is essentially a triangle with the slope of the sides matched to the gradual increase of the likelihood. Because the triangle is sloped everywhere, it also regularises in the vicinity of the true time delay. Ideally, we would like a triangle prior with a flat top, which would cancel the edge modes and acts as a uniform prior near the central mode. This begs the question in which region to flatten the triangle, which requires knowledge of where the true time delay is. However, this is what we intend to infer in the first place and hence do not know \emph{a priori}.

In any case, we generally do not advocate for biased priors to change pathological likelihood functions and instead seek changes in the method to sample unbiased yet pathological likelihoods.

As a final remark, we note that the regularised likelihood was constructed using knowledge about the functional form of the likelihood itself. From a Bayesian point of view, this may be an undesirable feature of a prior. On the other hand, this approach bears similarities to a Jeffreys prior~\cite{jeffreys1946invariant}, which is also constructed from the likelihood.

\subsection{Further considerations on the likelihood}\label{sec:synthetic-data-further-considerations}

In this section, we discuss caveats to the above arguments regarding the gradual drift of the log-likelihood.

\subsubsection{Likelihood or log-likelihoood?}

Firstly, we should actually consider the likelihood instead of the log-likelihood since 
by Jensen's inequality, $\mathbb{E}_\mathbf{y}L\ge \exp(\mathbb{E}_{\mathbf{y}}\log L)$, the data-averaged log-likelihood only provides a lower bound on the log of the data-averaged likelihood. Therefore, the data-averaged likelihood may be larger at any time delay, in particular at the boundary. This means that the likelihood at the boundary could in principle still lie above the likelihood at the true time delay, thus invalidating the above conclusions. As we show in Appendix~\ref{appendix:data-averaged-likelihood}, the data-averaged likelihood favours the true time delay, consistent with our conclusions. However, the variance diverges, so that it is not possible to draw conclusions which hold for any data set, in contrast to the data-averaged log-likelihood, for which the variance remains small. 
A similar divergence of the variance is displayed by the data-averaged Bayes factor
(Appendix~\ref{appendix:bayes-factor}).
Furthermore, the data-averaged likelihood, $\int\mathrm{d}\mathbf{y}\,p(\mathbf{y}\mid \bm{\theta})$, misses the normalisation of the posterior computed by nested sampling,
$\int\mathrm{d}\mathbf{y}\,\frac{p(\mathbf{y}\mid \bm{\theta})p(\bm{\theta})}{Z(\bm{y})}$, where $Z(\bm{y})$ denotes the Bayesian evidence. As we use uniform priors, the factor $p(\bm{\theta})$ can be neglected. At best, we can expect the factor $\frac{1}{Z(\mathbf{y})}$ to suppress the contribution to the posterior from data sets for which the peak at the true time delay is subdominant. 
However, we show numerically in Section~\ref{sec:numerical-problems} that this is not the case.

\subsubsection{Non-perturbative effects}

Secondly, the above derivation assumed that the noise~$\sigma$ is small, allowing a perturbative treatment. If $\sigma$ is comparable to $A$ or larger, $\mathbf{y}_1$ and $\mathbf{y}_2$ lose their mutual correlation, which does not align with the real data sets considered here. 
A more pressing concern, however, is that we fixed all (hyper)parameters to their true values except for the time delay. We address this question in Section~\ref{sec:synthetic-data-numerical-issues-2} below.

\subsubsection{Numerical conditioning}

Thirdly, we note that the covariance matrix used in the calculation of the likelihood can become rather ill-conditioned. This error propagates to the numerical value of the likelihood itself. We rule out a systematic influence of this error in Appendix~\ref{appendix:numerical-conditioning}, showing that the full covariance matrix $\mathbf{K}$ exhibits large condition numbers at regular values of $\Delta t$ but not at the edges. 
If the numerical conditioning did have an influence, it would be important to ensure that our analysis reflect the same numerical error.
Indeed, the calculation of the data-averaged log-likelihood (Equation~\ref{eqn:data-averaged-log-likelihood}) involves the inverse of $\mathbf{K}_{\bm{\theta}}$ which is implemented by a Cholesky decomposition, following the standard \gls{gp} implementation~\citep{rasmussen}.
In contrast, the data-averaged likelihood (Equation~\ref{eqn:data-averaged-likelihood}) involves the determinant of the matrix and is therefore not numerically representative of the log-likelihood used in \gls{ns} runs on real data. 

\subsubsection{Beyond stationarity}

Fourthly, if we allow for a non-stationary kernel, the error bars of the interpolation may not necessarily increase to their maximum value over the decorrelation length scale. More generally, non-stationary \glspl{gp} with non-constant mean functions can extrapolate trends and we could expect that the problem with the drift disappears. However, incorporating non-stationary \glspl{gp} in fits on real data~\cite{kroupa2026quasars} showed the problem of the large time delays persists, showing that it is more severe than anticipated and that the assumption of stationarity in this section can be relaxed. 

Finally, on real data, we do not know the data-generating distribution. Therefore, the \gls{gp} model may be misspecified. This certainly adds a further layer of complexity. Indeed, this may contribute to the fact that the large time delays occur empirically for real data.

\section{Numerical investigation}\label{sec:numerical-investigation}

\subsection{Method}

We sample the posterior with \gls{ns}, primarily for three reasons. Firstly, \gls{ns} naturally incorporates uniform priors and priors with hard constraints, such as those from the flexknots, as \gls{ns} can operate without gradients of the likelihood.
Secondly, the \gls{gp} likelihood with a joint uniform prior on all hyperparameters may include singular covariance matrices which lead to divergent negative log-likelihoods, i.e. zero likelihood. \gls{ns}, in particular \textsc{PolyChord}~\citep{handley2015polychord, handley}, adaptively ignores such regions without changing the evidence.
Thirdly, kernel, flexknot and time delay posteriors are observed to be typically multimodal and exhibit extended degeneracies, which global methods such as \gls{ns} handle automatically.

In practice, we use the \textsc{PolyChord} implementation of \gls{ns} and the \textsc{anesthetic} package~\citep{handley-anesthetic} for sample post-processing, unless stated otherwise.
All calculations are perfomed on the original \gls{gp} likelihood (Equation~\ref{eqn:synthetic-data-likelihood}) with a uniform priors, i.e. we do not regularise with the counterterm prior (Section~\ref{sec:counterterm-prior}).

\subsection{Sampling $\Delta t$}\label{sec:numerical-problems}

In this section, we create synthetic data sets of random \gls{gp} realisations with known time delay and infer the posterior time delay from the synthetic data.
We sample $\Delta t$ using \gls{ns} with a uniform prior on $[-t_\mathrm{range},t_\mathrm{range}]$, where $t_\mathrm{range}=T_\mathrm{max}-T_\mathrm{min}$. Default \textsc{PolyChord} settings are used. The sampling is performed for a data set drawn from Equation~\ref{eqn:time-delayed-GP-pair} with $\Delta t_\mathrm{true}=10\,\days$. This is repeated for $25$ independent and identically distributed data sets. 
We thus obtain a data-averaged posterior, formally defined as $\int \mathrm{d}\mathbf{y}\,p(\bm{\theta}\mid \mathbf{y})p(\mathbf{y}\mid \bm{\theta}_\mathrm{true})$, 
where 
$\bm{\theta}_\mathrm{true}$ are the parameters generating $\mathbf{y}$. 
In practice, this means that we merge the sets of equally-weighted posterior samples obtained from each data set and perform any further inference with this enlarged set of samples.

\begin{figure}
	\centering
	\includegraphics[width=\columnwidth]{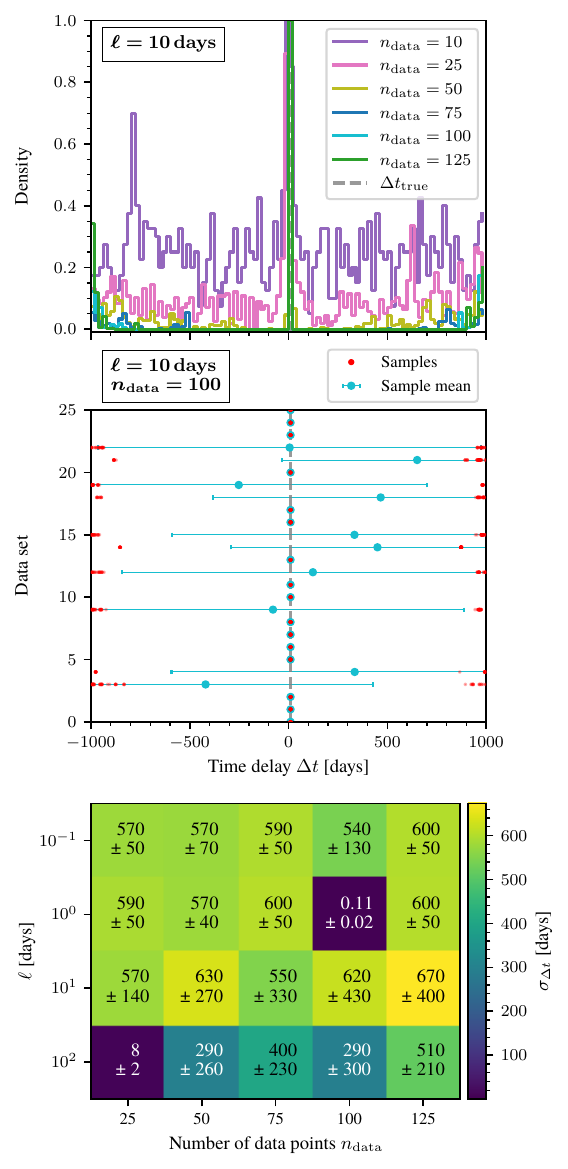}
	\caption{
		\textbf{Top:} Data-averaged posterior for $\ell=10\,\days$. As $n_\mathrm{data}$ is increased, the central peak of the posterior becomes sharper around the true time delay, as expected. However, the posterior density also increases around the edges, $\pm t_\mathrm{range}$. 
		\textbf{Middle:} Posterior time delays of each data set for $\ell=10\,\days$ and $n_\mathrm{data}=100$. The posterior is peaked at either the true time delay or $\pm t_\mathrm{range}$.
		\textbf{Bottom:} Standard deviation $\sigma_{\Delta t}$ of $\Delta t$ from the data-averaged posterior, for different values of $\ell$ and $n_\mathrm{data}$. The standard deviation is systematically around the same order of magnitude as $t_\mathrm{range}$, indicating that the inference of large time delays (as in the top subplot) are systematic across $\ell$ and $n_\mathrm{data}$. The error bars on each value indicate the variation of $\sigma_{\Delta t}$ between different data set realisations. Evidently, $\sigma_{\Delta t}$ can vary significantly between data sets.
	}
	\label{fig:synthetic-data-delta-t}
\end{figure}

The results are shown in Figure~\ref{fig:synthetic-data-delta-t}. The top subplot shows the data-averaged posterior for $\ell=10\,\days$. As the number of observations $n_\mathrm{data}$ is increased, the uncertainty in the posterior is suppressed and the peak at the true time delay becomes increasingly sharp. This is consistent with the expectation that more data should improve the inference. 
More precisely, the data density on the time axis increases as $n_\mathrm{data}$ is increased. At the critical value of $n_\mathrm{data}=100$, there is a data point every $\frac{t_\mathrm{range}}{n_\mathrm{data}}=10\,\days=\ell$, i.e. samples start to become correlated. This correlation structure in the data is expected to improve time delay inference. Stated differently, the data points are effectively uncorrelated for $n<100$ so that both light curves appear as two uncorrelated random draws so that there is non-zero probability density away from the true time delay.

Counterintuitively, in addition to the peak around the true value, the posterior becomes sharper around the edges of the distribution, i.e. at $\pm t_\mathrm{range}$, as $n_\mathrm{data}$ is increased.
To obtain a more refined view of the posterior, the middle subplot shows the decomposition of the data-averaged posterior into the contribution from individual data sets for $n_\mathrm{data}=100$. The posterior is peaked at either the true time delay, in which case the error bar on $\Delta t$ is small, or at both edges, $\pm t_\mathrm{range}$, in which case the error bar on ${\Delta t}$ is large. 

It remains to ask whether these large time delays consistently appear across different values of $\ell$. To quantify the width of the data-averaged posterior, we compute the standard deviation $\sigma_{\Delta t}$ of $\Delta t$. In agreement with the above description of the data-averaged posterior, the bottom subplot shows that $\sigma_{\Delta t}$ remains roughly unchanged for $\ell=10\,\days$ as $n_\mathrm{data}$ is increased. 
For $\ell<10\,\days$, $\sigma_{\Delta t}$ remains large but tends to become smaller for $\ell>10\,\days$. This is consistent with the above argument that correlation between data points improve time delay inference.
However, $\sigma_{\Delta t}$ tends to increase with $n_\mathrm{data}$ for $\ell=100\,\days$ and $\ell=125\,\days$ as more data points increase the concentration of the posterior at the edges. 
Note that the fluctuations between the data sets also remain large since $\sigma_{\Delta t}$ is an unstable estimator and hence is expected to converge very slowly with the number of data sets.
Overall, the above results suggest that, depending on the randomly drawn data set, the inferred time delay either agrees with the true time delay or is significantly far away. This observation is systematic across different (hyper)parameters of our setup.

Increasing the number of live points by a factor of $10$ fixes the problem by replacing the inferred large time delays with the true time delay.
Figure~\ref{fig:data-averaged-loglikelihood} provides the underlying reason. A \gls{ns} run is initialised by populating the $\Delta t$-axis uniformly with live points. The probability of a live point landing in the sharp mode at the true time delay depends on the prior volume of that mode and becomes smaller as the number of data points is increased. Suppose that no live point lands in the mode. As the \gls{ns} run then proceeds, the live points are pushed up in likelihood while retaining a uniform distribution. The probability of a live point hopping into the mode decreases as the \gls{ns} iteration number increases, partly because the prior volume at the current likelihood contour decreases. But more importantly, the gradual drift in the log-likelihood (Section~\ref{sec:log-likelihood-structure}) pushes the live points away from the mode, towards the edges $\Delta t=\pm t_\mathrm{range}$, so that they aggregate in the local modes at the edges. This explains why \gls{ns} either infers the true time delay (if a live point happens to land in the true mode at initialisation) or a large time delay (if no live points lands in that mode and they are pushed away from it). Moreover, this also explains why the convergence failure becomes more extreme when the number of data points increases. 
Consequently, increasing the number of live points increases the probability of finding the true mode at the initialisation of a \gls{ns} run as the resolution of the parameter space is increased.

For real data sets and \gls{gp} models with more parameters, this problem becomes more severe since the fractional parameter space volume of the true mode is even smaller.

\subsection{Convergence by number of live points}\label{sec:convergence-by-number-of-live-points}

\begin{figure}
	\centering
	\includegraphics{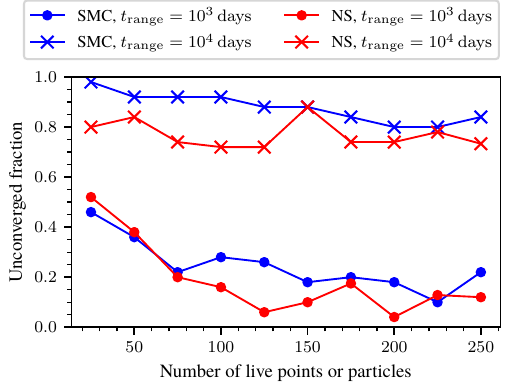}
	\caption{
		Fraction of unconverged Nested Sampling (NS) and Sequential Monte Carlo (SMC) runs against number of live points and particles, respectively. As the number is increased, non-convergence becomes less probable since finding the central mode becomes more probable. Thus, the fraction decreases. 
		Increasing the observation window $t_\mathrm{range}$ from $10^3\,\days$ to $10^4\,\days$ increases the probability of non-convergence as the central mode occupies a smaller fraction of parameter space and the gradual rise of the likelihood towards the edge modes drives the population of live points or particles further away from the central mode.
		Similar trends are seen for both NS and SMC.
		The plotted lines are only visual guides.
	}
	\label{fig:quantitative-convergence}
\end{figure}

To make the previous point about increasing the number of live points to ensure convergence more precise, we perform 
$50$ independent and identically distributed \gls{ns} runs in the setting of Section~\ref{sec:numerical-problems} for each value of the number of live points. For each ensemble of runs, we compute the fraction of unconverged runs. 

An unconverged run is defined as follows. Given a likelihood (induced by a randomly drawn dataset), we perform a Laplace approximation at the local model corresponding to the true time delay. The location of the mode is obtained by local optimisation, initialised at the true time delay which generated the data, and the standard deviation around this mode is obtained via the second derivative.
A \gls{ns} run is said to be unconverged if the posterior mean of the time delay is more than a certain number $f$ of standard deviations away from the Laplace approximation. 
Note that the standard deviation used here is the one computed from the Laplace approximation. We do not use the \gls{ns} posterior standard deviation, as this can be very large and falsely indicate convergence.
As the Laplace approximation is only a good approximation when the number of observations is large, we expect that even a converged \gls{ns} posterior may not be in good agreement (i.e. within, say, $3$ standard deviations) with the Laplace approximation. 
We must therefore calibrate $f$ to reflect the level of agreement defining convergence. For this, we increase the number of live points until we determine visual agreement between the two time delay posteriors (as in the previous sections) for multiple \gls{ns} runs. We then read off the maximum disagreement measured in standard deviations and set $f$ to this value. Hence, for this large value of the live points, there are by definition no unconverged runs. 

To reduce the computational cost of the \gls{ns} runs, we use the \textsc{BlackJAX} implementation of \gls{ns}~\cite{yallup2025nested, yallup2026nested} on a GPU, which internally uses slice sampling like \textsc{PolyChord}. We choose the same termination criterion as in \textsc{PolyChord} and, like $\textsc{PolyChord}$, the number of slice sampling steps to be $5$ times the number of dimensions, which is $1$ since we are only sampling the time delay. These settings make the calculations in this section comparable to the previous ones. The Laplace approximation is implemented via automatic differentiation in \textsc{jax}~\cite{frostig}. The local optimisation is performed via Newton's method~\cite{press2007numerical}.

Figure~\ref{fig:quantitative-convergence} shows the fraction of unconverged runs against the number of live points, for light curves with $100$ observations and an observation windows $t_\mathrm{range}$ of $10^3\,\days$ and $10^4\,\days$. 
Here we discuss the results for \gls{ns}. The results for Sequential Monte Carlo are discussed in Section~\ref{sec:other-global-samplers-numerical-example}.
As expected, for $10^3\,\days$, the fraction decreases as the number of live points is increased. This is because the probability of initialising a live point in the central mode increases with the number of live points. We see that even at $250$ live points, there is still a non-zero probability of missing the central mode. Note that $250$ live points is an order of magnitude larger than the standard setting adopted by \textsc{PolyChord}.
To show that the pathology is driven by the relative volumes of the central mode to the edge modes, as well as the gradual rise which guides live points slowly away from the central mode during a \gls{ns} run, we increase the observation window of the light curves to $10^4\,\days$. The fraction of unconverged runs increases for all considered values of live points. While a slow decrease in the fraction is visible, consistent with our expectation, this implies that the number of live points must be substantially increased to ensure proper convergence.

Finally, we note that the calculations in Figure~\ref{fig:quantitative-convergence} are visibly noisy, since the fraction does not decrease monotonically. The noise can be decreased further by increasing the number of independent \gls{ns} runs or improving the quality of the Laplace approximation by considering datasets with substantially more observations. Both of these options become quickly infeasible due to the computational costs.

\subsection{Sampling $\Delta t$, $\ell$ and $\sigma$}\label{sec:synthetic-data-numerical-issues-2}

Here, we address the question of whether the structure of the log-likelihood is preserved if the length scale $\ell$ and noise $\sigma$ are sampled as well.
For each of these hyperparameter values, the full joint posterior over $\bm{\theta}=(\Delta t, \ell, \sigma)$ is sampled with \gls{ns} with standard settings. The priors are set to be uniform in the ranges $\ell\in [0, t_\mathrm{range}]$, $\Delta t\in [-t_\mathrm{range},t_\mathrm{range}]$ and $\sigma\in [0,1]$, where $t_\mathrm{range}$ is the difference between the maximum and minimum observation time. 
The parameters are set to $\ell=10\,\days$, $\Delta t=10\,\days$ and $n_\mathrm{data}\in\{10, 100, 200, 400, 600, 800\}$. Due to the increased computational cost, we perform the inference only on $12$ independent and identically distributed data sets.

We observe that the inferred $\sigma$ is larger than the true value and the posterior modes at the edges disappear, leaving just the mode around the true value of $\Delta t$. This is consistent with the caveat discussed in Section~\ref{sec:synthetic-data-further-considerations}. However, due to the small sample size of $12$ data sets, we cannot exclude the possibility that there exist data sets for which the edge modes are not suppressed.

Instead, we show that the likelihood still exhibits a ``W''-shape by reducing the convergence parameters on our \gls{ns} runs. 
Specifically, the \textsc{PolyChord} settings \texttt{num\_repeats} and \texttt{nlive} are now set to $3$ and $15$ instead of the default values $15$ and $75$, respectively. 
Lowering \texttt{nlive} has the effect of coarse-graining the parameter space so that the \gls{ns} run is more susceptible to missing modes with small parameter space volume. Lowering \texttt{num\_repeats} induces correlations between \gls{ns} dead points and makes the set of live points susceptible to following funnel-shaped likelihoods during the \gls{ns} run. This allows us to investigate the structure of the likelihood surface in statistically less significant regions (which a proper \gls{ns} run would ignore) while circumventing a direct plot of the now higher-dimensional log-likelihood. We refrain from plotting the log-likelihood directly as such plots are visually dominated by numerically highly unstable, albeit statistically negligible, low log-likelihood regions.

\begin{figure}[!]
	\centering
	\includegraphics{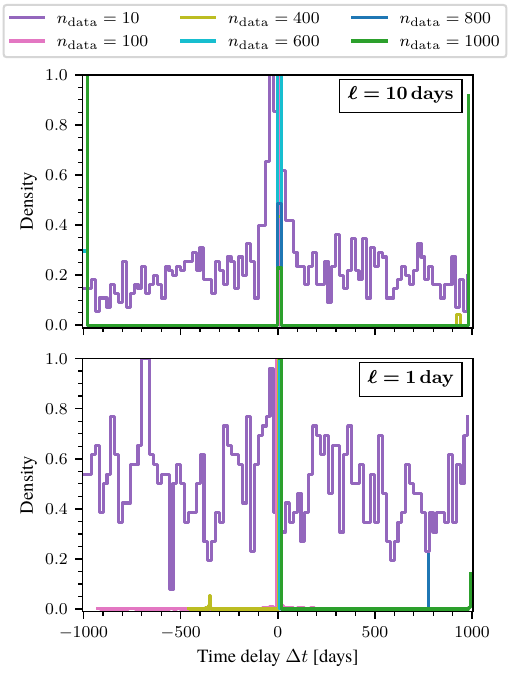}
	\caption{
		Data-averaged posteriors with reduced nested sampling convergence settings. \textbf{Top}: The data-averaged posterior of the time delay converges to a trimodal distribution as the density of data points increases: While the central peak at the true time delay is visible, two unphysical additional modes at the edges of the observation time interval emerge. \textbf{Bottom}: The unphysical modes are suppressed by increasing the effective sample size of the \gls{gp}. This is achieved by decreasing the length scale by a factor of $10$. In both figures, the inference was performed on synthetic light curves generated with known time delay $\Delta t_\mathrm{true}$ and \gls{gp} length scale $\ell$.}
	\label{fig:synthetic-data}
\end{figure}

The top subplot of Figure~\ref{fig:synthetic-data} shows the data-averaged time delay posterior for $\Delta t=10\,\days$ and $\ell=10\,\days$ and $1\,\mathrm{day}$. For $n_\mathrm{data}=10$ data points, the posterior is noisy and the uncertainty around $\Delta t_\mathrm{true}$ is correspondingly large. As $n_\mathrm{data}$ is further increased, a sharp peak at $\Delta t_\mathrm{true}$ develops. However, two additional modes at the edges of the observation time, $\pm 10^3\,\days$, emerge. Moreover, if we inspect the posterior for each data set individually, we observe that each posterior is peaked at exactly one of the three modes of the data-averaged posterior. 

This means that, for some data sets, the inference of $\Delta t_\mathrm{true}$ is possible, whereas it is not possible for other data sets. Since all data sets are random draws from the same \gls{gp} distribution, this implies that the random fluctuations in a particular \gls{gp} realisation control whether the true time delay can be inferred. In any case, the underlying structure of the log-likelihood still exhibits the drift to the edges.

We note that the synthetic light curves have fewer effective samples than the number of data points suggests as the observations are correlated with a decorrelation length scale $\ell_\mathrm{true}$. We can thus define the effective sample size as $\frac{T}{\ell_\mathrm{true}}$, where $T$ is the total observation time. Hence, increasing the number of effective samples further can be achieved by decreasing $\ell$. 
The bottom subplot of Figure~\ref{fig:synthetic-data} shows the same calculation but with $\ell_\mathrm{true}$ decreased by a factor of $10$.
This partially suppresses the modes at the edges of the data-averaged posterior, consistent with the expectation that more independent data points should improve the time delay inference, however further spikes appear away from the true time delay. 

\section{Real data and cosmologial priors}

In this section, we move beyond toy problems and investigate the likelihood induced by real data and whether priors from cosmology and lens models are sufficient to alleviate the problems presented previously.

\subsection{Real light curve data}

\begin{figure}[!]
	\centering
	\includegraphics{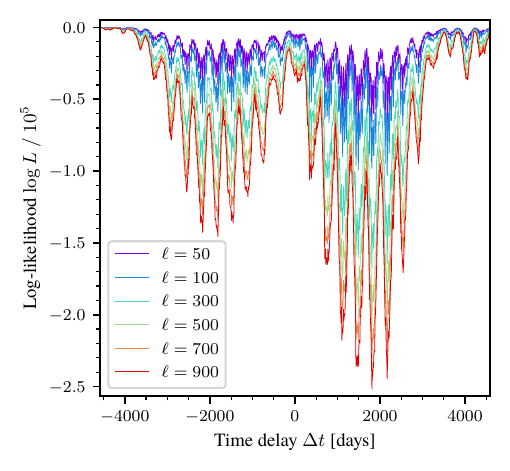}
	\caption{
		Cuts through the log-likelihood surface for light curve data from the lensed quasar \he. The log-likelihood exhibits a ``W''-shape, with the value of $\log L$ at the edges comparable to $\log L$ near the central peak. This drives nested sampling towards the edges of the time delay prior, explaining systematic convergence problems~\cite{kroupa2026quasars}.}
	\label{fig:real-data-cut-through-logL}
\end{figure}

The numerical investigation so far has been limited to synthetic data from a toy model. Time delay inference calculations~\cite{kroupa2026quasars} showed that this problem persists on real light curve data under a broad choice of mean functions and kernels. 

To show that the log-likelihood still exhibits the global structure described previously, we plot an exemplary cut through the high-dimensional log-likelhood surface for a pair of light curves of the lensed quasar \he. While the data-generating \gls{gp} is unknown in this case, we choose the exponential kernel and remove the choice of the mean function and amplitude hyperparameter by whitening the pair of light curves, i.e. for each light curve, we subtract the mean and divide the magnitudes by the standard deviation. We then plot the log-likelihood as a function of the time delay for different kernel length scales $\ell$, as shown in Figure~\ref{fig:real-data-cut-through-logL}.
Manifestly, the log-likelihood qualitatively shows the same global structure. Moreover, the values of the log-likelihood at the edges of the time delay prior are comparable to those at the central peak, which we associate with the true time delay, i.e. the globally dominant mode of the posterior. Note that the likelihood being comparable or higher at the edges than the one at the centre is not an issue in a fully Bayesian framework as the parameter space volume of each mode must also be taken into account.

\subsection{Time delay priors from lens models and cosmology}

The priors considered so far are based solely on the light curves. Cosmological parameter inference additionally requires a prior on the cosmology and the lens model. We may therefore ask whether the cosmological and lens model priors induce a sufficiently constrained time delay prior which excludes the preference for large time delays.

For this, we take a simplified version of a lens model for \mbox{\he} described in~\citet{wong2017h0licow} as an example. We adopt a singular power-law elliptical mass distribution~\citep{barkana1999fastell} for the lens galaxy, combined with an external shear component. This model includes the following free parameters: the Einstein radius $\theta_\mathrm{E}\in[0,100]^{\prime\prime}$, power-law slope $\gamma\in[0,100]$, ellipticity components $e_1\in[-\frac12,\frac12]$ and $e_2\in[-\frac12,\frac12]$, smoothing scale $s_\mathrm{scale}\in[0,100]^{\prime\prime}$, and lens centroid position ${\theta_{x,y}\in[-2,2]^{\prime\prime}}$. The external shear is parameterised by its two components $\gamma_{1,2}\in[-\frac12,\frac12]$. 
We fix the lens and source redshifts to $z_\mathrm{d} = 0.4546$ \citep{morgan2005lens} and $z_\mathrm{s} = 1.693$ \citep{sluse2012microlensing}, respectively.
For the cosmological priors, we adopt a flat $\Lambda$CDM model with uniform distributions for the Hubble parameter ${H_0 \in [0, 100]\, \mathrm{km}\,\mathrm{s}^{-1}\,\mathrm{Mpc}^{-1}}$ and matter density ${\Omega_\mathrm{m} \in [0, 1]}$. 

We generate samples of lens configurations by drawing parameter values from uniform prior distributions over their respective ranges. A configuration is accepted if exactly four images are formed from a central source position $(0^{\prime\prime},0^{\prime\prime})$, consistent with the observed quad configuration in \he. 
The time delays $\Delta t$ are computed by taking pairwise differences between the arrival times of the images. The arrival times are calculated using the \texttt{lenstronomy} package \citep{birrer2018lenstronomy, birrer2021lenstronomy}. We also show the results for the same calculations performed for the narrower prior ${H_0 \in [60, 80]\, \mathrm{km}\,\mathrm{s}^{-1}\,\mathrm{Mpc}^{-1}}$ and discuss this further below.

\begin{figure*}[!]
	\centering
	\includegraphics{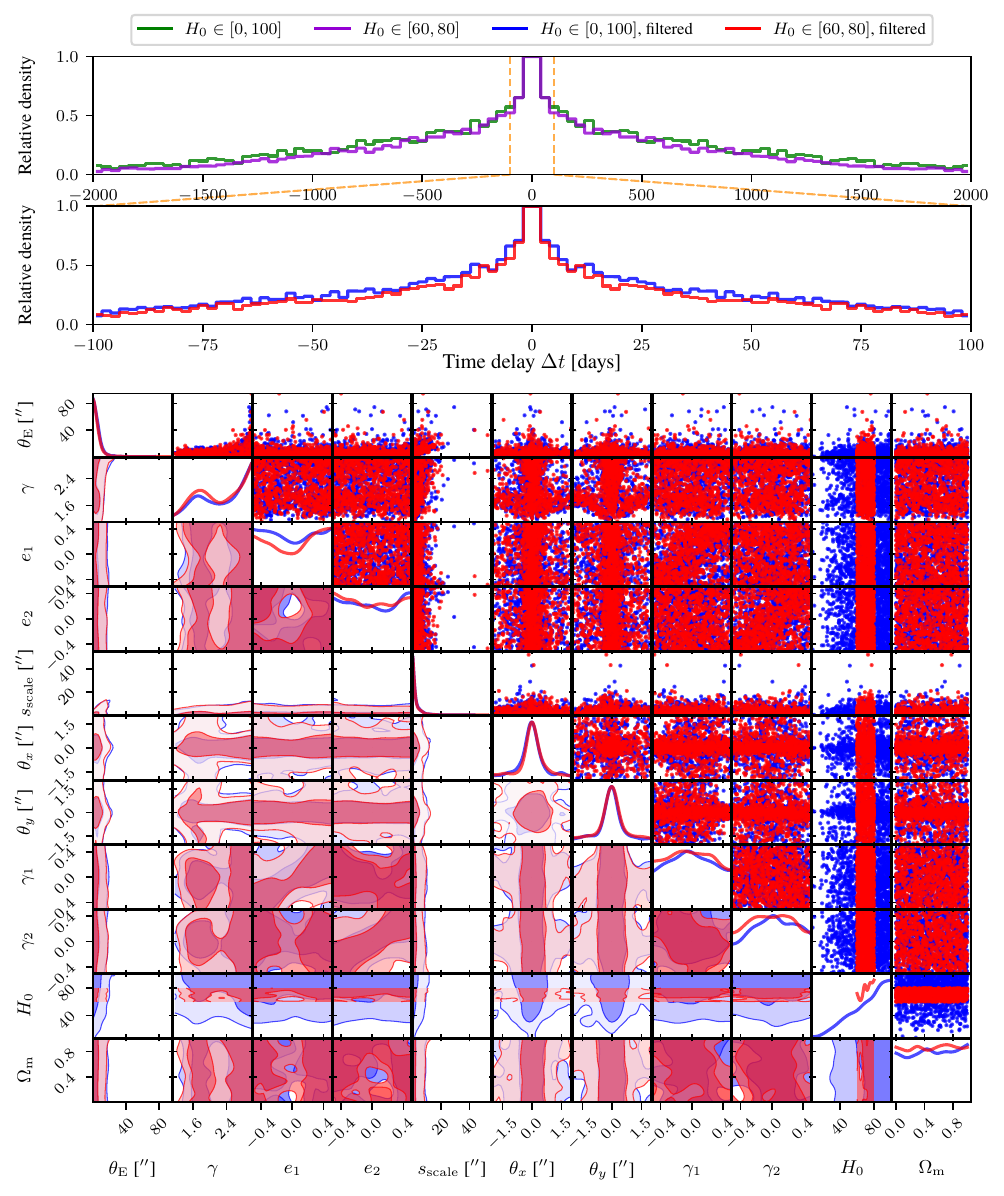}
	\caption{\textbf{Top:} Histogram of time delays induced by uniform lens model and cosmology priors, filtering for four images. Larger time delays are visibly penalised but there is significant probability mass at $\pm 2000\days$. Hence, these priors do not sufficiently constrain the time delay. Narrowing the prior of $H_0$ to the range $[60,80]$ induces a narrower distribution on the time delay, as evidenced by the lighter tails of the time delay distribution. This effect is negligible, however. \textbf{Middle:} Same histogram, but with time delays filtered to $\pm 100\,\days$. \textbf{Bottom:} Corresponding distributions on the lens model and cosmology parameters induced by the filtering to $\pm 100\,\days$. These distributions show the priors which would have given rise to the restricted time delays. Note that $H_0$ acquires a bias to larger values. Restricting the $H_0$ prior range to $[60,80]$ again barely changes the distributions.}
	\label{fig:time-delay-histogram}
\end{figure*}

The resulting time delays are collected in a histogram, shown in the top plot of Figure~\ref{fig:time-delay-histogram}. We observe that a significant fraction of the probability mass lies in the range $[0,2000]\,\days$. Hence, we do not expect the lens model and cosmological priors to constrain the time delays sufficiently for time delay inference. 

However, we indeed observe that the prior density decreases as $|\Delta t|$ increases. 
This shows that large time delays, i.e. of the order of $10^3\,\days$ for this system, are unphysical, as previously anticipated.
The shape of the distribution is expected since the lens model and geometry of the system only constrain the product $H_0 \Delta t$~\citep{narayan1996lectures}, so that a uniform prior on $H_0$ induces a distribution on $\Delta t$ which diverges at $\Delta t=0$. Sampling lens models and the cosmologies amounts to forming a weighted superposition these divergent distributions, leading to the observed shape. 

As discussed in the Introduction,  we assess the impact of setting an implicit prior on the time delay by repeating the above calculation of the histogram but retaining only those configurations that produce time delays within $\pm 100\,\days$.
The resulting histogram is shown in the middle plot of Figure~\ref{fig:time-delay-histogram}. This is the restriction of the histogram in the top plot to the desired range. We additionally show the distribution of the lens model and cosmology parameters induced by this filtering in the bottom plot. These distributions show which priors we would have had to set in order to produce the restricted time delay distribution.

As follows from the above discussion, we expect that small values of $H_0$ are removed by removing large time delays from the sample set. This is confirmed by the distribution shown in the corner plot, in which $H_0$ acquires an approximately linear bias to larger values. This bias propagates to posteriors when performing inference and we leave an analysis of implications thereof to further work.
In contrast, the prior on $\Omega_\mathrm{m}$ remains roughly uniform.

Following this result, we could attempt to narrow the range of the $H_0$ prior in order to constrain the $\Delta t$ prior.
The results are also shown in Figure~\ref{fig:time-delay-histogram}.
Narrowing the range to $[60, 80]\, \mathrm{km}\,\mathrm{s}^{-1}\,\mathrm{Mpc}^{-1}$ still produces time delays of order $2000\,\days$ and thus does not sufficiently constrain the time delay. The tails of the time delay distributions become lighter, although this difference is negligible.
We again show the filtered prior samples in Figure~\ref{fig:time-delay-histogram} and the corresponding corner plot, which reinforce that the narrow prior has negligible influence on the previous conclusions.

We remark that the distributions shown in the corner plot are manifestly non-uniform. For all lens model parameters, most of the non-uniformity comes from conditioning on the quad configuration of \he. For the cosmological parameters, this conditioning has negligble influence and most of the non-uniformity is a result of restricting the time delay prior range.
This strongly suggests that the lens model parameters are primarily constrained by imaging data, not the light curves.

\section{Other global samplers}

In this section, we discuss other global sampling algorithms, namely \gls{pt}~\cite{hukushima1996exchange, earl2005parallel} and \gls{smc}~\cite{chopin2002sequential, del2006sequential} and diffusive~\gls{ns}~\cite{brewer2011diffusive}.

\subsection{Discussion}

\gls{pt} and \gls{smc} are tempering methods, which sample a sequence of $K$ intermediate distributions 
$p_{\beta_i}\propto \pi L^{\beta_i}$
with $0=\beta_1<\beta_2<\cdots<\beta_K=1$ between prior and posterior corresponding to $\beta_1=0$ and $\beta_K=1$, respectively.

\gls{pt} maintains such a ladder of temperatures and runs \gls{mcmc} at each temperature, proposing and accepting swaps between temperatures.
Hot chains ($\beta\ll 1$) cross barriers and swaps exchange information between hot and cold chains ($\beta\sim 1$).
In the case of time delay inference, the true mode is extremely narrow in prior volume and there is a broad, gently rising region towards the edges which provides moderately large likelihood over a large prior volume. \gls{pt} may still struggle because the probability a chain ever visits a region with tiny volume can still be vanishingly small, even if barriers are flattened.

Moreover, \gls{pt} relies on sufficient overlap between the distributions $p_{\beta_i}$ of adjacent temperatures as this controls the swap acceptance rate. If a cold distribution has a very narrow spike at the true delay but hotter distributions are dominated by the edges, then adjacent \gls{mcmc} chains can have little overlap in log-likelihood. This requires many temperatures, $K\gg 1$, in order to increase the overlap and long mixing at each temperature, which increases the computational cost in a way analogous to increasing the number of live points in the \gls{ns} setting.

\gls{smc} maintains a distribution of particles starting at the prior and evolves them simultaneously from hot to cold distributions. It does this by reweighting and resampling particles, and subsequently mutation via \gls{mcmc}. 
If the central mode has a tiny measure under intermediate $\beta$ (i.e. the probability of finding a particle in the central mode is tiny), while the boundary region has a huge measure, then before rejuvenation is able to discover the central mode, resampling preferentially resamples boundary particles. After a few iterations through different values of $\beta$, the population becomes dominated by boundary ancestors.
This problem is resolved by sufficiently increasing the number of particles.
Overall, this is analogous to the problem of \gls{ns}, in which the central mode is missed at initialisation, and where the number of particles serves an analogous role as the number of live points.

Diffusive \gls{ns} is a variation of \gls{ns} which aims to improve exploration of the parameter space by sampling a mixture over likelihood levels, in a fashion which is similar to simulated annealing over the constrained \gls{ns} level sets.
However, the failure mode in the case of time delay inference is that the algorithm never sees the true basin in the first place. Once the level structure and the \gls{mcmc} moves are built from samples which mostly live near the edges, we still end up with levels which are constructed from the edge modes and \gls{mcmc} moves within a level which rarely hit the tiny central mode. 
That is, diffusive \gls{ns} does not remove the need for at least one particle or trajectory to enter the tiny central mode at some point.
This problem is again resolved by increasing the number of levels or equivalently increasing the number of live points.

In summary, these global sampling algorithms, including \gls{ns}, are built on the same principle, namely by bridging a unimodal and easy-to-sample prior and a multimodal ergodically-broken posterior. If the spurious boundary modes posed local traps, then these methods make it possible to escape them. In the setting of time delay inference, the edge modes are not traps but instead supported by a global trend and large phase space volume. The bridging makes the distribution more volume-dominated, so the fundamental problem is finding and retaining the true mode.

\subsection{Numerical example}\label{sec:other-global-samplers-numerical-example}

To illustrate the preceding discussion, we provide a numerical example with \gls{smc} with an identical setup as in Section~\ref{sec:convergence-by-number-of-live-points}. 
Importantly, we observed that the fraction of unconverged runs was positive for all tested hyperparameter settings of \gls{smc}. As we are interested in the behaviour of the fraction as the number of particles is increased, in particular its decay as the number of particles is increased, the specific value of the fraction is less important. 
Thererfore, the hyperparameters of \gls{smc} are chosen to roughly match the fraction of unconverged runs of \gls{ns} at $25$ particles and live points, respectively.

Specifically, we use an adaptive tempered \gls{smc} sampler implemented in \textsc{BlackJAX}~\cite{cabezas2024blackjax}.
For each run, a fixed number of particles are drawn from the prior. At each iteration $i$, the next temperature $\beta_{i+1}$ is selected automatically to achieve a target relative effective sample size of $0.5$. 
A root-finding procedure via bisection is used to determine the increment $\Delta\beta=\beta_{i+1}-\beta_i$.
Particles are reweighted by the incremental likelihood power with weights $w_{i+1}\propto w_i L^{\Delta\beta}$, followed by resampling to control weight degeneracy. 
After resampling, particles are mutated using a random-walk Metropolis \gls{mcmc} chain targeting the current tempered distribution $\pi_{\beta_i}$. We use a Gaussian proposal with mean zero and covariance adapted at each \gls{smc} stage from the empirical weighted particle covariance, as done in the \textsc{BlackJAX} implementation of \gls{ns}~\cite{yallup2025nested} which we are comparing with in Section~\ref{sec:convergence-by-number-of-live-points}. 
The covariance is scaled by a heuristic value of $0.125$.
The algorithm iterates until $\beta=1$ is reached. 
A final particle approximation of the posterior is given by standard \gls{smc} estimators.

Figure~\ref{fig:quantitative-convergence} shows the convergence results for \gls{smc}. The decreasing trend of the fraction of unconverged runs with increasing particle number is shared with \gls{ns}, as expected. Similarly, increasing the length of the observation window $t_\mathrm{range}$ increases the fraction as particles are more likely to be driven to the edges. This increase is roughly of the same size as for \gls{ns}, pointing to the fact that the pathology is intrinsic to the structure of the time delay likelihood and independent of the sampling algorithm.

\section{Conclusions}\label{sec:conclusions}

The time delay likelihood possesses a fundamental pathological structure, appearing as a ``W''-shaped landscape with a dominant central peak at the true value and spurious modes rising at the boundaries of the observation window. Denser sampling over a fixed time span sharpens both the true peak and these edge modes. This structure is not a shortcoming of a particular machine learning model but instead a fundamental consequence of time delay inference itself. In particular, because a light curve is shifted out of the region of overlapping data, the problem is intrinsically extrapolative. \glspl{gp} make this issue explicit, as they revert to their prior away from conditioning data and must effectively extrapolate a random walk process. However, the effect is model-independent, as any finite observation window necessitates extrapolation.

In principle, the inference problem is well-posed, as the peak at the true delay dominates when sampled correctly. The practical difficulty, however, lies in the sampling methodology. Global methods like nested sampling require stringent settings, including substantially more live points, tighter termination criteria, multiple seeds, and stability checks, to overcome this structure of the likelihood. Otherwise, the live points tend to drift along the boundary induced likelihood gradient and fail to resolve the central peak. Conversely, local methods and optimisers that implicitly favour small values of $|\Delta t|$ can bias $H_0$ 
to larger values. While external priors from lens models and cosmology rarely suppress the large delay modes sufficiently, informative, system-specific priors can offer partial mitigation, though they do not represent a universal solution.

Beyond astrophysics, the problem analysed in this work is at the centre of scalability issues of \gls{ns} in materials science~\cite{chew2023phase}, as explained in the introduction.
In particular, this has seeded efforts for the development of novel extensions to \gls{ns}~\cite{martiniani2014superposition, baldock2016determining, unglert2025replica}.
In contrast to the materials case, however, where the dimension of the sampling space is often of order $10^2$ to $10^3$, with the goal of letting the dimensionality become as large as possible to approach the thermodynamic limit, the time delay problem is low-dimensional as it scales with the number of gravitationally lensed images, and therefore vastly more interpretable.
Novel sampling strategies designed to alleviate this difficulty may therefore be carefully tested and analysed for the time delay problem, which is significantly more difficult in the case of materials science. 
Further, a solution to one problem may imply a solution for the other.
Together with recent advances in understanding concerned with the challenge of incorporating gradient information into \gls{ns}~\cite{kroupa2025resonances}, our work continues to push the frontier of \gls{ns} method development.

From an algorithmic perspective, \gls{ns} may be viewed as a particular variation of \gls{smc}~\cite{salomone2018unbiased}, which constitutes a much broader design space of global sampling algorithms. Again, a solution to the time delay problem for \gls{ns} may imply a transferable and wider-reaching solution in the larger class of \gls{smc} algorithms, thus rendering the time delay problem a testbed for novel global sampling algorithms.

Overall, we envision that our work provides a foundation for developing further solutions to this sampling instability. Coupled with continued algorithmic development, these efforts will pave the way for fully Bayesian time delay inference, light curve regression and beyond.

\section*{Acknowledgements}
N. K. thanks G\'abor Cs\'anyi for helpful discussions.
N. K. was supported by the Harding Distinguished Postgraduate Scholarship. 
This work was performed using the Cambridge Service for Data Driven Discovery (CSD3), part of which is operated by the University of Cambridge Research Computing on behalf of the STFC DiRAC HPC Facility (www.dirac.ac.uk). The DiRAC component of CSD3 was funded by BEIS capital funding via STFC capital grants ST/P002307/1 and ST/R002452/1 and STFC operations grant ST/R00689X/1. DiRAC is part of the National e-Infrastructure.

\appendix

\section{Derivation for the time delay posterior}\label{appendix:derivation-time-delay-posterior}

Here, we give a derivation of Equation~\ref{eqn:time-delay-posterior}. By Bayes' theorem, we have
\begin{equation}
	p(\Delta t\mid y_1,y_2)=\frac{p(y_1,y_2\mid\Delta t)p(\Delta t)}{p(y_1,y_2)},
\end{equation}
with prior $p(\Delta t)$, likelihood $p(y_1,y_2\mid\Delta t)$ and evidence $p(y_1,y_2)=p(y_2\mid y_1)p(y_1)$. The likelihood may be written in the following marginalised form:
\begin{align}
	p(y_1,y_2\mid\Delta t)
	&=\int\mathrm{d}\mathbf{w}\,p(y_1,y_2,\mathbf{w}\mid\Delta t)\\
	&=\int\mathrm{d}\mathbf{w}\,p(y_2\mid \Delta t,y_1,\mathbf{w})p(y_1,\mathbf{w}\mid\Delta t),
\end{align}
where the integration is performed over all parameters ${\mathbf{w}=(w_1,w_2,\dots)^\top}$.
The last term decomposes further as
\begin{equation}
	p(y_1,\mathbf{w}\mid\Delta t)=p(\mathbf{w}\mid y_1,\Delta t)p(y_1\mid \Delta t),
\end{equation}
Here, $p(y_1\mid \Delta t)$ is the prior probability of $y_1$ given $\Delta t$. However, $y_1$ only depends on $\Delta t$ if we are given $y_2$. Therefore, we may simplify this as $p(y_1\mid \Delta t)=p(y_1)$.
Further, $p(\mathbf{w}\mid y_1,\Delta t)$ is the posterior of $\mathbf{w}$ given only the first light curve $y_1$. This is independent of $\Delta t$ and hence simplifies to $p(\mathbf{w}\mid y_1)$, as well.

Considering $p(y_2\mid \Delta t,y_1,\mathbf{w})$, we note that $y_2$ is completely determined by $\Delta t$ and $\mathbf{w}$ and independent of $y_1$ as $\Delta t$ and $\mathbf{w}$ determine the underlying function $f$. In particular, we may write this as
\begin{equation}
	p(y_2\mid \Delta t,y_1,\mathbf{w})=\prod_{i=1}^{n_\mathrm{data}}\delta (y_2(t_i)-f(t_i+\Delta t,\mathbf{w})),
\end{equation} 
where $\delta$ is the Dirac delta function, but we choose to keep this term in the more general form $p(y_2\mid \Delta t,\mathbf{w})$.

Collecting the above equations, we are left with
\begin{equation}
	p(\Delta t\mid y_1,y_2)=\frac{p(\Delta t)}{p(y_2\mid y_1)}\int\mathrm{d}\mathbf{w}\,p(y_2\mid \Delta t,\mathbf{w})p(\mathbf{w}\mid y_1).
\end{equation}
We usually choose the prior $p(\Delta t)$ to be uniform so that the prefactor of the integral is a constant. This leads to Equation~\ref{eqn:time-delay-posterior}, as required.

\section{Weakly determined length scales}\label{appendix:weakly-determined-length-scales}

We consistently observe that the length scale posterior of the exponential kernel increases for small values of the length scale and subsequently
exhibits 
a degenerate flat posterior on the scale of the time range of the data set, up to the end of the prior range. In contrast, the other kernels have a much more constrained posterior. It is therefore reasonable to ask whether the problem lies with the data or with the exponential kernel itself. In the following, we conclude that it must be the data and that the posterior is in principle not pathological under optimal assumptions.

For a wide uniform prior, the posterior is proportional to the likelihood. Hence we only consider the likelihood from here on. We assume that the given data $\mathbf{y}$ is sampled in regular intervals $\delta t$ from a zero-mean \gls{gp} with the true exponential kernel $\bm{\Sigma}_{ij}=b^{|i-j|}$ with unit amplitude and true length scale $\ell_\mathrm{true}$, where $b=\mathrm{e}^{-\delta t/\ell_\mathrm{true}}$. 
The log-likelihood is then
\begin{equation}\label{eqn:weakly-determined-length-scales-logL}
	\log L(\ell)=-\frac12 \mathbf{y}^\top\mathbf{K}^{-1}\mathbf{y}-\frac12\log |\mathbf{K}|-\frac{n}{2}\log(2\pi),
\end{equation}
where $n$ is the total number of data points (i.e. number of light curves times the number of data points $n_\mathrm{data}$ per light curve), the covariance matrix $\mathbf{K}_{ij}=a^{|i-j|}$ with $a=\mathrm{e}^{-\delta t/\ell}$ depends on the parameter $\ell$ and we have set the amplitude in the covariance matrix to its true value for simplicity. We have $\mathbf{y}=\mathbf{K}^{1/2}\hat{\mathbf{y}}$, where $\hat{\mathbf{y}}$ is a sample from a Gaussian with an identity covariance matrix. We now consider the large data limit, in which $n\rightarrow \infty$. In this limit, the individual terms in Equation~\ref{eqn:weakly-determined-length-scales-logL} become self-averaging and we may use random matrix theory~\cite{tulino2004random, tao2012topics, potters2020first} to proceed further. To this end, we introduce the normalised trace operator
\begin{equation}
	\tau(\mathbf{A})=\lim_{n\rightarrow\infty}\frac{1}{n}\mathbb{E}[\mathrm{Tr}\mathbf{A}],
\end{equation}
for an $n\times n$ matrix $\mathbf{A}$. 

Using the cyclic property of the trace, the first term in Equation~\ref{eqn:weakly-determined-length-scales-logL} can now be written as
\begin{align}
	-\frac12 \mathbf{y}^\top\mathbf{K}^{-1}\mathbf{y}
	&=
	-\frac{n}{2}\tau({\bm{\Sigma}^{1/2}\mathbf{K}^{-1}\bm{\Sigma}^{1/2}\hat{\mathbf{y}}\hat{\mathbf{y}}^\top})\\
	&=-\frac{n}{2}\tau({\bm{\Sigma}^{1/2}\mathbf{K}^{-1}\bm{\Sigma}^{1/2}})\tau(\hat{\mathbf{y}}\hat{\mathbf{y}}^\top)
\end{align}
In the last line we have used the fact that $\tau$ factorises to leading order for the free product of a deterministic and random matrix. Notably, $\tau(\hat{\mathbf{y}}\hat{\mathbf{y}}^\top)=1$ since $\hat{\mathbf{y}}\hat{\mathbf{y}}^\top$ is a Wishart matrix. Using the cyclic property of the trace again, we have $\tau({\bm{\Sigma}^{1/2}\mathbf{K}^{-1}\bm{\Sigma}^{1/2}})=\tau(\mathbf{K}^{-1}\bm{\Sigma})$. The inverse $\mathbf{K}^{-1}$ is known analytically,
\begin{equation}
	(\mathbf{K}^{-1})_{ij} = \frac{1}{1-a^2} \begin{cases}
		1 & \text{if } i=j=1 \text{ or } i=j=n, \\
		1+a^2 & \text{if } 1 < i=j < n, \\
		-a & \text{if } |i-j|=1, \\
		0 & \text{else},
	\end{cases}
\end{equation}
from which we obtain $\tau(\mathbf{K}^{-1}\bm{\Sigma})=\frac{1+a^2-2ab}{1-a^2}$.

To handle the second term in Equation~\ref{eqn:weakly-determined-length-scales-logL}, we can approximate $\mathbf{K}$ by a circulant matrix since boundary effects become negligible in the large data limit,
\begin{align}
	\mathbf{K}&=\begin{bmatrix}
		1 & a & \cdots & a^{n-1}\\
		a & 1 & \cdots & a^{n-2}\\
		\vdots & \vdots & \ddots & \vdots\\
		a^{n-1} & a^{n-2} & \cdots & 1\\
	\end{bmatrix}\\
	&\approx\begin{bmatrix}
		1 & a & a^2 & \cdots & a^2 & a\\
		a & 1 & a   & \cdots & a^3 & a^2\\
		a^2 & a & 1 & \cdots & a^4 & a^3 \\
		\vdots & \vdots & \vdots & \ddots & \vdots & \vdots\\
		a & a^2 & a^3 & \cdots & a & 1\\
	\end{bmatrix}.
\end{align}
This is equivalent to considering a \gls{gp} on a circle with periodic boundary conditions and letting the radius become large.

The spectrum is then given by 
\begin{equation}
	\lambda(x_i)=\frac{1-a^2}{1+a^2-2a\cos(\pi x_i)},
\end{equation}
where $x_i=2i/n$ with $i\in\{0,\dots,n/2\}$. Using this, we have
\begin{align}
	-\frac12\log |\mathbf{K}|&=-\frac12 \mathrm{Tr}\log\mathbf{K}\\
	&=-\frac{n}{2} \tau(\log\mathbf{K})\\
	&=-\frac{n}{2}\int_{0}^1\log\lambda(x)\mathrm{d}x\\
	&=-\frac{n}{2}\log(1-a^2).\label{eqn:e-kernel-log-det}
\end{align}
where in the third line the sum over eigenvalues was replaced by an integral and in the final line $\int_0^1\log(1+a^2-2a\cos(\pi x))\mathrm{d}x=0$ for all $a\in [0,1]$ was used.

Therefore, the log-likelihood per data point now becomes
\begin{equation}
	\frac{\log L(\ell)}{n}=-\frac{1}{2} \left(\frac{1+a^2-2 a b}{1-a^2}+\log \left(1-a^2\right)+\log (2 \pi ) \right).
\end{equation}
Maximising this log-likelihood by setting the derivative to zero yields $\ell=\ell_\mathrm{true}$, as desired.

We can now investigate the large $\ell$ limit. To leading order in $\ell$, we have $\log L(\ell)/n\approx -\frac{1}{2}\frac{1-b}{\delta t}\ell$. Thus, $\log L(\ell)\rightarrow -\infty$ linearly in $\ell$. Hence, we likelihood is well constrained and we do not expect any degeneracies under optimal data assumptions.

\section{A refined analysis of the log-likelihood}\label{appendix:refined-argument-log-likelihood}

Here, we offer more clarity on the intuitive description of the log-likelihood given in Section~\ref{sec:log-likelihood-structure}.

First, we rewrite Equation~\ref{eqn:time-delayed-GP-pair} in the following sequential form:
\begin{align}
	\mathbf{y}_1\sim \mathcal{N}(&\mathbf{0},\mathbf{K}(\mathbf{t},\mathbf{t})+\sigma^2\mathbf{I})\\
	\mathbf{y}_2\mid \mathbf{y}_1\sim \mathcal{N}(&\mathbf{m}_{\mathbf{y}_2\mid \mathbf{y}_1},\mathbf{K}_{\mathbf{y}_2\mid\mathbf{y}_1})\label{eqn:factorised-GP-eqn2},
\end{align}
where 
\begin{align}
	\mathbf{m}_{\mathbf{y}_2\mid \mathbf{y}_1}&=\mathbf{K}(\mathbf{t}-\Delta t\mathbf{1},\mathbf{t})[\mathbf{K}(\mathbf{t},\mathbf{t})+\sigma^2\mathbf{I}]^{-1}\mathbf{y}_1,\\
	\mathbf{K}_{\mathbf{y}_2\mid\mathbf{y}_1}&=\mathbf{K}(\mathbf{t},\mathbf{t})+\sigma^2\mathbf{I}\\
	&\phantom{=}-\mathbf{K}(\mathbf{t}-\Delta t\mathbf{1},\mathbf{t})[\mathbf{K}(\mathbf{t},\mathbf{t})+\sigma^2\mathbf{I}]^{-1}\mathbf{K}(\mathbf{t}-\Delta t\mathbf{1},\mathbf{t})\notag
\end{align}
The conditional form of a multivariate normal distribution was used here, see \citet{rasmussen}, Equation~A.6. 
The interpretation of Equation~\ref{eqn:factorised-GP-eqn2} is that $\mathbf{y}_1$ was interpolated, giving the smooth interpolant $\mathbf{m}_{\mathbf{y}_2\mid \mathbf{y}_1}$ with error bars $\mathbf{K}_{\mathbf{y}_2\mid\mathbf{y}_1}$, and we look at the interpolant at the shifted times $\mathbf{t}-\Delta t \mathbf{1}$. Thus, we pretend that the data $\mathbf{y}_2$ has observation times $\mathbf{t}-\Delta t\mathbf{1}$ and compare this to the interpolant. Also, note that $\mathbf{m}_{\mathbf{y}_2\mid \mathbf{y}_1}$ is a linear combination of kernels centered on each data point $\mathbf{y}_1$.

The likelihood correspondingly factorises, 
\begin{equation}
	p(\mathbf{y}_1,\mathbf{y}_2\mid \Delta t)=p(\mathbf{y}_2\mid\mathbf{y}_1, \Delta t)p(\mathbf{y}_1\mid \Delta t),
\end{equation}
which states that we first compare $\mathbf{y}_1$ with the \gls{gp} prior, i.e. we evaluate $p(\mathbf{y}_1\mid \Delta t)$ at the data set $\mathbf{y}_1$. Then we interpolate $\mathbf{y}_1$, shift the data set $\mathbf{y}_2$ to the times $\mathbf{t}-\Delta t \mathbf{1}$ and compare with the interpolant by evaluating ${p(\mathbf{y}_2\mid\mathbf{y}_1, \Delta t)}$. Since the term ${p(\mathbf{y}_1\mid \Delta t)}$ actually does not depend on $\Delta t$ and therefore only contributes a constant factor to the likelihood, we focus on the behaviour of $p(\mathbf{y}_2\mid\mathbf{y}_1, \Delta t)$. This log of this term can be decomposed into a $\chi^2$ term measuring the goodness of fit and a penalty term:
\begin{align}
	&\log p(\mathbf{y}_2\mid\mathbf{y}_1, \Delta t)\\&=-\frac12 \underbrace{(\mathbf{y}_2-\mathbf{m}_{\mathbf{y_2\mid\mathbf{y}_1}})^\top \mathbf{K}_{\mathbf{y}_2\mid\mathbf{y}_1}^{-1}(\mathbf{y}_2-\mathbf{m}_{\mathbf{y_2\mid\mathbf{y}_1}})}_{\chi^2}\notag\\
	&\phantom{=}-\frac12\underbrace{\log\left|\mathbf{K}_{\mathbf{y}_2\mid\mathbf{y}_1}\right|}_{\text{penalty}}-\frac{n}{2}\log(2\pi),\notag
\end{align}
where we recall that $n$ is the total number of data points, i.e. the number of light curves times the number of data points $n_\mathrm{data}$ per light curve.

Suppose now that the true time delay is $\Delta t_\mathrm{true}=0$, as this simplifies the analysis, and that the noise is negligible compared to the kernel amplitude, $\sigma\ll A$.
We can then consider the interpolated mean and covariance at the true time delay, $\Delta t=0$, and for large time delays, meaning that~$\Delta t$ is much larger than the length scale $\ell$ of the kernel. 

For $\Delta t=0$, we have to first order in $\sigma$,
\begin{align}
	\mathbf{m}_{\mathbf{y}_2\mid\mathbf{y}_1}&\approx \mathbf{y}_1-\sigma^2\mathbf{K}(\mathbf{t},\mathbf{t})^{-1}\mathbf{y}_1,\\
	\mathbf{K}_{\mathbf{y}_2\mid\mathbf{y}_1}&\approx 2\sigma^2\mathbf{I},
\end{align}
meaning that $\mathbf{y}_2$ is modelled as a copy of $\mathbf{y}_1$ with a small amount of distortion of order $\mathcal{O}(\sigma^2)$ and random noise of order $\mathcal{O}(\sigma)$.
Using $\mathbf{K}_{\mathbf{y}_2\mid\mathbf{y}_1}\approx \mathcal{O}(\sigma^2)$, it can be shown that $\mathbf{y}_2-\mathbf{y}_1=\mathcal{O}(\sigma)$. Thus, we have
\begin{equation}
	\mathbf{y}_2-\mathbf{m}_{\mathbf{y}_2\mid\mathbf{y}_1}=\mathcal{O}(\sigma)+\sigma^2\mathbf{K}(\mathbf{t},\mathbf{t})^{-1}\mathbf{y}_1=\mathcal{O}(\sigma)
\end{equation}
and hence $\chi^2=\mathcal{O}(n)$ to leading order in $\sigma$. The penalty term is $\mathcal{O}(n\log(2\sigma^2))$ which is large and negative. Therefore, the penalty dominates and we observe a large and positive value of $\log p(\mathbf{y}_2\mid\mathbf{y}_1, \Delta t)$, leading to the peak at the true time delay.

For $\Delta t$ large, we first note that $\mathbf{K}(\mathbf{t}-\Delta t\mathbf{1},\mathbf{t})\approx \mathbf{0}$ since the kernel is stationary. Hence, we immediately obtain
\begin{align}
\mathbf{m}_{\mathbf{y}_2\mid\mathbf{y}_1}&\approx \mathbf{0},\\
\mathbf{K}_{\mathbf{y}_2\mid\mathbf{y}_1}&\approx \mathbf{K}(\mathbf{t},\mathbf{t})+\sigma^2\mathbf{I},
\end{align}
meaning that the light curves are modelled as decorrelated from each other. Since the (unconditional) distribution of $\mathbf{y}_2$ is $\mathcal{N}(\mathbf{0},\mathbf{K}(\mathbf{t},\mathbf{t})+\sigma^2\mathbf{I})$, we have that $\chi^2=\mathcal{O}(n)$. 

Further, the penalty term is $\log\left|\mathbf{K}(\mathbf{t},\mathbf{t})+\sigma^2\mathbf{I}\right|\approx\log\left|\mathbf{K}(\mathbf{t},\mathbf{t})\right|$, which in general is a function of $A$, $\ell$ and $n$. 
For example for the exponential kernel, comparison with the result from Equation~\ref{eqn:e-kernel-log-det} gives $\log|\mathbf{K}(\mathbf{t},\mathbf{t})|= n(2\log A+\log(1-\mathrm{e}^{-2\delta t/\ell}))$ in the large data limit, where we had to pull the factor of $A^2$ out of the matrix to use the result, which holds for $A=1$. Since $\log(1-\mathrm{e}^{-2\delta t/\ell})\ge 0$, we have that $\log|\mathbf{K}(\mathbf{t},\mathbf{t})|\ge 2n\log A = \mathcal{O}(n\log A)$.
We assume that a similar scaling holds for other stationary kernels. The important point is that the penalty can be taken to be of order $\mathcal{O}(n\log A)$.

We can now collect the above results and compare the size of terms. For $\Delta t=0$ and large $\Delta t$, the penalty term dominates with values $\mathcal{O}(n\log(2\sigma^2))$ and $\mathcal{O}(n\log A)$, respectively. Since $\sigma\ll A$, the former value is much more negative than the latter and this is what ultimately favours the time delay $\Delta t=0$.
For both values of the time delay, we have comparable values of $\chi^2=\mathcal{O}(n)$. This is subdominant compared to the penalty term.

\section{Data-averaged likelihood}\label{appendix:data-averaged-likelihood}

The integral of the likelihood (Equation~\ref{eqn:synthetic-data-likelihood}) with $\mathbf{y}$ distributed according to Equation~\ref{eqn:time-delayed-GP-pair} gives directly:
\begin{align}
	\mathbb{E}_\mathbf{y}L(\bm{\theta})&=|2\pi(\mathbf{K}_{\bm{\theta}}+\mathbf{K})|^{-1/2},\label{eqn:data-averaged-likelihood}\\
	\mathbb{E}_\mathbf{y}L^2(\bm{\theta})&=|4\pi^2\mathbf{K}_{\bm{\theta}}(\mathbf{K}_{\bm{\theta}}+2\mathbf{K})|^{-1/2},
\end{align}
where $\mathbf{K}_{\bm{\theta}}$ and $\mathbf{K}$ are $n\times n$ matrices. Here, $n$ is the total number of data points, i.e. number of light curves times the number of data points $n_\mathrm{data}$ per light curve.

Taking the true time delay to be zero, we have at ${\Delta t=0}$, $\mathbf{K}_{\bm{\theta}}=\mathbf{K}$. Hence,
\begin{align}
	\mathbb{E}_\mathbf{y}L(\bm{\theta})&=(4\pi)^{-n/2}|\mathbf{K}|^{-1/2}\\
	&=(4\pi)^{-n/2}|\mathbf{K}(\mathbf{t},\mathbf{t})+\sigma^2\mathbf{I}|^{-1/2}\\
	\times|\mathbf{K}(\mathbf{t},\mathbf{t})&+\sigma^2\mathbf{I}-\mathbf{K}(\mathbf{t},\mathbf{t})(\mathbf{K}(\mathbf{t},\mathbf{t})+\sigma^2\mathbf{I})^{-1}\mathbf{K}(\mathbf{t},\mathbf{t})|^{-1/2}\\
	&=(4\pi)^{-n/2}\sigma^{-n}|2\mathbf{K}(\mathbf{t},\mathbf{t})+\sigma^2\mathbf{I}|^{-1/2}
\end{align}
where we have used the formula for the determinant of a block matrix and the fact that $\mathbf{K}(\mathbf{t},\mathbf{t})+\sigma^2\mathbf{I}$ and $\mathbf{K}(\mathbf{t},\mathbf{t})$ commute. Similarly,
\begin{equation}
	\mathbb{E}_\mathbf{y}L^2(\bm{\theta})=(12\pi^2)^{-n/2}\sigma^{-2n}|2\mathbf{K}(\mathbf{t},\mathbf{t})+\sigma^2\mathbf{I}|^{-1}.
\end{equation}
Thus, the expectation and standard deviation both scale as $\mathcal{O}(\sigma^{-n})$ and diverge for small $\sigma$.

For $\Delta t$ large, 
\begin{equation}
	\mathbf{K}_{\bm{\theta}}=\begin{bmatrix}
		\mathbf{K}(\mathbf{t},\mathbf{t})+\sigma^2\mathbf{I}&\mathbf{0}\\
		\mathbf{0}&\mathbf{K}(\mathbf{t},\mathbf{t})+\sigma^2\mathbf{I}
	\end{bmatrix}.
\end{equation}
To calculate $|\mathbf{K}_{\bm{\theta}}+\mathbf{K}|$, we again use the determinant of a block matrix. This requires the use of Jacobi's formula,
\begin{equation}
|\mathbf{K}(\mathbf{t},\mathbf{t})+\sigma^2\mathbf{I}|\approx |\mathbf{K}(\mathbf{t},\mathbf{t})|(1+\sigma^2\mathrm{Tr}[\mathbf{K}(\mathbf{t},\mathbf{t})^{-1}]),
\end{equation}
from which we obtain $|\mathbf{K}_{\bm{\theta}}+\mathbf{K}|\approx 3^{n}|\mathbf{K}(\mathbf{t},\mathbf{t})|^2$ to leading order in $\sigma$. This gives us
\begin{equation}
	\mathbb{E}_\mathbf{y}L(\bm{\theta})\approx(6\pi)^{-n/2}|\mathbf{K}(\mathbf{t},\mathbf{t})|^{-1}.
\end{equation}
By a similar procedure, we obtain
\begin{equation}
	\mathbb{E}_\mathbf{y}L^2(\bm{\theta})\approx (20\pi^2)^{-n/2}|\mathbf{K}(\mathbf{t},\mathbf{t})|^{-2}.
\end{equation}
Thus, the expectation and standard deviation remain finite for small $\sigma$ and large $\Delta t$.

\section{Bayes factor}\label{appendix:bayes-factor}

The Bayes factor between the true time delay $\Delta t=0$ and large time delay, $\Delta t=\infty$, is given by
\begin{align}
	B&=\frac{p(\mathbf{y}\mid \Delta t=0)}{p(\mathbf{y}\mid \Delta t=\infty)}\\
	&=\sqrt{\frac{|\mathbf{K}_\infty|}{|\mathbf{K}_0|}}\exp\left(-\frac{1}{2}\mathbf{y}^\top(\mathbf{K}_0^{-1}-\mathbf{K}_\infty^{-1})\mathbf{y}\right),
\end{align}
where $\mathbf{K}_0$ and $\mathbf{K}_\infty$ are $n\times n$ matrices.
To calculate the variance, $\mathbb{E}_\mathbf{y}B^2$ must be finite. We now show that this is not the case.

Since $\mathbf{y}$ is distributed as $\mathcal{N}(\mathbf{0},\mathbf{K}_0)$, the Gaussian integral $\mathbb{E}_\mathbf{y}B^2$ is finite if the matrix $\mathbf{Q}=3\mathbf{K}_0^{-1}-2\mathbf{K}_\infty^{-1}$ is positive definite. This is equivalent to showing that, for every non-zero $\mathbf{v}$, $\mathbf{v}^\top\mathbf{Q}\mathbf{v}>0$. Noting that $\mathbf{K}_0$ is positive definite, we change basis to $\mathbf{w}=\mathbf{K}_0^{-1/2}\mathbf{v}$ so that the condition becomes $\mathbf{w}^\top\mathbf{K}_0^{-1/2}\mathbf{S}\mathbf{K}_0^{1/2}\mathbf{w}>0$, where we defined $\mathbf{S}=3\mathbf{I}-2\mathbf{K}_0\mathbf{K}_\infty^{-1}$. This is a similarity transform of $\mathbf{S}$ by $\mathbf{K}_0^{-1/2}$. Since a similarity transform preserves positive definiteness, it remains to show that $\mathbf{S}$ is positive definite. 

We now calculate the eigenvalues of $\mathbf{S}$.
We have
\begin{align}
	&\mathbf{K}_0\mathbf{K}_\infty^{-1}
	=\begin{bmatrix}
		\mathbf{K}(\mathbf{t},\mathbf{t})+\sigma^2\mathbf{I}&\mathbf{K}(\mathbf{t},\mathbf{t})\\
		\mathbf{K}(\mathbf{t},\mathbf{t})&\mathbf{K}(\mathbf{t},\mathbf{t})+\sigma^2\mathbf{I}
	\end{bmatrix}\\
	&\phantom{=}\times 
	\begin{bmatrix}
		(\mathbf{K}(\mathbf{t},\mathbf{t})+\sigma^2\mathbf{I})^{-1}&\mathbf{0}\\
		\mathbf{0}&(\mathbf{K}(\mathbf{t},\mathbf{t})+\sigma^2\mathbf{I})^{-1}
	\end{bmatrix}\\
	&=\begin{bmatrix}
		\mathbf{I}&\mathbf{K}(\mathbf{t},\mathbf{t})(\mathbf{K}(\mathbf{t},\mathbf{t})+\sigma^2\mathbf{I})^{-1}\\
		\mathbf{K}(\mathbf{t},\mathbf{t})(\mathbf{K}(\mathbf{t},\mathbf{t})+\sigma^2\mathbf{I})^{-1}&\mathbf{I}
	\end{bmatrix}
\end{align}
Since for any block symmetric matrix, it holds that
\begin{equation}
\begin{vmatrix}\mathbf{A}&\mathbf{B}\\\mathbf{B}&\mathbf{A}\end{vmatrix}=\left|\mathbf{A}+\mathbf{B}\right| \left|\mathbf{A}-\mathbf{B}\right|,
\end{equation}
the eigenvalues of $\mathbf{K}_0\mathbf{K}_\infty^{-1}$ are those of ${\mathbf{I}+\mathbf{K}(\mathbf{t},\mathbf{t})(\mathbf{K}(\mathbf{t},\mathbf{t})+\sigma^2\mathbf{I})^{-1}}$ and ${\mathbf{I}-\mathbf{K}(\mathbf{t},\mathbf{t})(\mathbf{K}(\mathbf{t},\mathbf{t})+\sigma^2\mathbf{I})^{-1}}$. Let $\kappa_i$, $i\in\{1,\dots,n/2\}$, be the eigenvalues of $\mathbf{K}(\mathbf{t},\mathbf{t})$. Then $\mathbf{K}_0\mathbf{K}_\infty^{-1}$ has eigenvalues 
$\lambda_i=1+\frac{\kappa_i}{\kappa_i+\sigma^2}$ for $i\in\{1,\dots,n/2\}$ and $\lambda_i=1-\frac{\kappa_i}{\kappa_i+\sigma^2}$ for $i\in\{n/2+1,\dots,n\}$. Hence, $\mathbf{S}$ has eigenvalues $\rho_i=3-2\lambda_i$. Expanding to leading order in $\sigma$, these are
\begin{equation}
	\rho_i\approx 
	\begin{cases}
	-1+2\frac{\sigma^2}{\kappa_i}&\text{for }i\in\{1,\dots,n/2\},\\
	3-2\frac{\sigma^2}{\kappa_i}&\text{for }i\in\{n/2+1,\dots,n\}.
	\end{cases}
\end{equation}

Following the argument in Appendix~\ref{appendix:refined-argument-log-likelihood}, the eigenvalues of $\kappa_i$ are of order $\mathcal{O}(n\log A)$ and thus non-negligible. Therefore, there are $\rho_i<0$ for small $\sigma$, i.e. $\mathbf{S}$ and hence $\mathbf{Q}$ is not positive definite so that the original integral under consideration diverges.

\section{Numerical conditioning}\label{appendix:numerical-conditioning}

\begin{figure}
	\centering
	\includegraphics{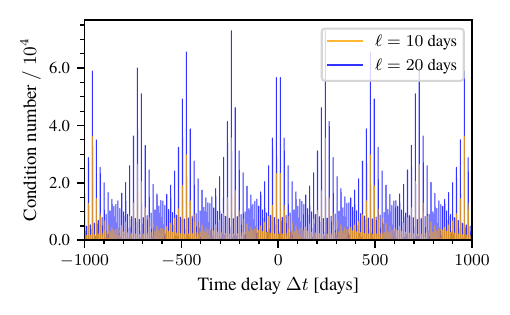}
	\caption{
		Condition number of the \gls{gp} covariance matrix. The plot was created with $6\times 10^4$ evenly spaced points along the time delay axis.}
	\label{fig:condition-number}
\end{figure}

We define the condition number of a covariance matrix as the ratio of the largest and smallest eigenvalues. Importantly, a larger condition number corresponds to a more singular matrix. The numerical errors accumulated in matrix operations, such as a Cholesky decomposition, are also correspondingly larger.

Figure~\ref{fig:condition-number} shows the condition number of the \gls{gp} covariance matrix,
\begin{equation}
	\begin{bmatrix}
		\mathbf{K}(\mathbf{t},\mathbf{t})+\sigma^2\mathbf{I}&\mathbf{K}(\mathbf{t},\mathbf{t}-\Delta t\mathbf{1})\\
		\mathbf{K}(\mathbf{t}-\Delta t\mathbf{1},\mathbf{t})&\mathbf{K}(\mathbf{t},\mathbf{t})+\sigma^2\mathbf{I}
	\end{bmatrix},
\end{equation}
as a function of $\Delta t$, with fixed $\sigma=0.01$ and $100$ evenly spaced points between $0$ and $10^3$ in $\mathbf{t}$. The condition number spikes at regular intervals. This is expected since some of the values in $\mathbf{t}$ and $\mathbf{t}-\Delta t\mathbf{1}$ are identical for certain values of $\Delta t$, which corresponds to two data points measured at the same time. At these values of $\Delta t$, the condition number is regularised by the $\sigma^2 \mathbf{I}$ terms. Indeed, increasing $\sigma$ reduces the height of the spikes. Further, the Figure shows that increasing $\ell$ also increases condition numbers. As we have more data points within the same correlation length, this is consistent with the previous result.

Notably, we do not observe exceptionally large condition numbers at the edges of the time delay prior. Hence, we conclude that the increase in the likelihood at the edges of the time delay prior does not stem from numerical ill-conditioning.

\bibliography{quasar_ml_time_delays}

\end{document}